\newcommand{\hst}{{\emph{HST}}}

\newcommand\Msun{\hbox{M$_\odot$}}
\newcommand\Zsun{\hbox{Z$_\odot$}}
\newcommand\kms{\hbox{$\,$km$\,$s$^{-1}$}}

\newcommand\one{\,{\sc i}}

\newcommand\three{\,{\sc iii}}

\newcommand\eg{e.\,g.}

\newcommand\bb{$B_{438}$}
\newcommand\vb{$V_{606}$}
\newcommand\ib{$I_{814}$}

\newcommand{\altcite}{\citealt}

\newcommand{\captionfonts}{\footnotesize}
\makeatletter  % Allow the use of @ in command names
\long\def\@makecaption#1#2{%
  \vskip\abovecaptionskip
  \sbox\@tempboxa{{\captionfonts #1: #2}}%
  \ifdim \wd\@tempboxa >\hsize
    {\captionfonts #1: #2\par}
  \else
    \hbox to\hsize{\hfil\box\@tempboxa\hfil}%
  \fi
  \vskip\belowcaptionskip}
\makeatother   % Cancel the effect of \makeatletter
\documentclass[12pt,natbib209,preprint]{aastex}
\usepackage{natbib}
\usepackage{times}
\usepackage{graphicx}
\usepackage{color}
\usepackage{epsf}
\usepackage{subfigure}
\renewcommand{\baselinestretch}{\spacing}
\shorttitle{History of interactions in SQ}
\shortauthors{K. Fedotov et al.}
\begin{document}
\title{Star Clusters as Tracers of Interactions in Stephan's Quintet (Hickson Compact Group 92)$^{\ast}$}
\author{K.~Fedotov\altaffilmark{1},
S.~C. Gallagher\altaffilmark{1},
I.~S.~Konstantopoulos\altaffilmark{2},
R.~Chandar\altaffilmark{3},
N.~Bastian\altaffilmark{4},
J.~C. Charlton\altaffilmark{2}
B.~Whitmore\altaffilmark{5},
\& 
G.~Trancho\altaffilmark{6}
}
\altaffiltext{*}{Based on observations made with the NASA/ESA Hubble Space Telescope.}
\altaffiltext{1}{Department of Physics \& Astronomy, The University of
  Western Ontario, London, ON, N6A 3K7, CANADA} 
\altaffiltext{2}{Department of Astronomy and Astrophysics, The Pennsylvania State University, University Park, PA, 16802, USA}  
\altaffiltext{3}{University of Toledo, Toledo, OH, 43606, USA} 
\altaffiltext{4}{School of Physics, University of Exeter, Stocker Road, Exeter, EX4 4QL, UK}
\altaffiltext{5}{Space Telescope Science Institute, Baltimore, MD, 21218, USA}
\altaffiltext{6}{Gemini Observatory, Casilla 603, La Serena, Chile}
\clearpage

\begin{abstract}
Stephan's Quintet (SQ; also known as Hickson Compact Group 92) is a compact group of galaxies that exhibits numerous signs of interactions between its members. Using high resolution (0\farcs04 per pixel) images of Stephan's Quintet in \bb, \vb, and \ib\ bands from the Early Release Science project obtained with the Wide Field Camera 3 on the \emph{Hubble Space Telescope}, we identify 496 star cluster candidates (SCCs), located throughout the galaxies themselves as well as in intergalactic regions. Our photometry goes $\sim 2$ mag deeper and covers an additional three regions, the Old Tail, NGC~7317, and the Southern Debris Region, compared to previous work. Through comparison of the \bb\ $-$ \vb\ and \vb\ $-$ \ib\ colors of the star cluster candidates with simple stellar population synthesis models we are able to constrain cluster ages. In particular,  the most massive galaxy of SQ, NGC~7319, exhibits continuous star formation throughout its history, although at a lower rate over the past few tens of Myr. NGC~7318 A/B and the Northern Star Burst region both show ongoing active star formation; there are a number of star clusters that are younger than $10$ Myr. NGC~7318 A/B also features a peculiar gap in the color distribution of the star clusters that can be used to date the onset of the recent burst. The majority of the SCCs detected in the Young Tail were formed $150$---$200$ Myr ago whereas the tight distribution of star cluster colors in the Old Tail, allow us to constrain its age of formation to $\sim 400$ Myr ago. The star clusters in the Southern Debris region are seemingly divided into two groups with ages of $50$ and $\sim 500$ Myr and virtually all of the SCCs detected in NGC~7317 are over $2$ Gyr old. Based on these ages, we estimate time intervals for the interactions between Stephan's Quintet members that triggered the massive star cluster formation.
\end{abstract}

%=============== SECTION 1 ===============
\section{Introduction}\label{sec:intro}
\begin{figure}
\begin{center}
	\includegraphics[width=1.0\textwidth]{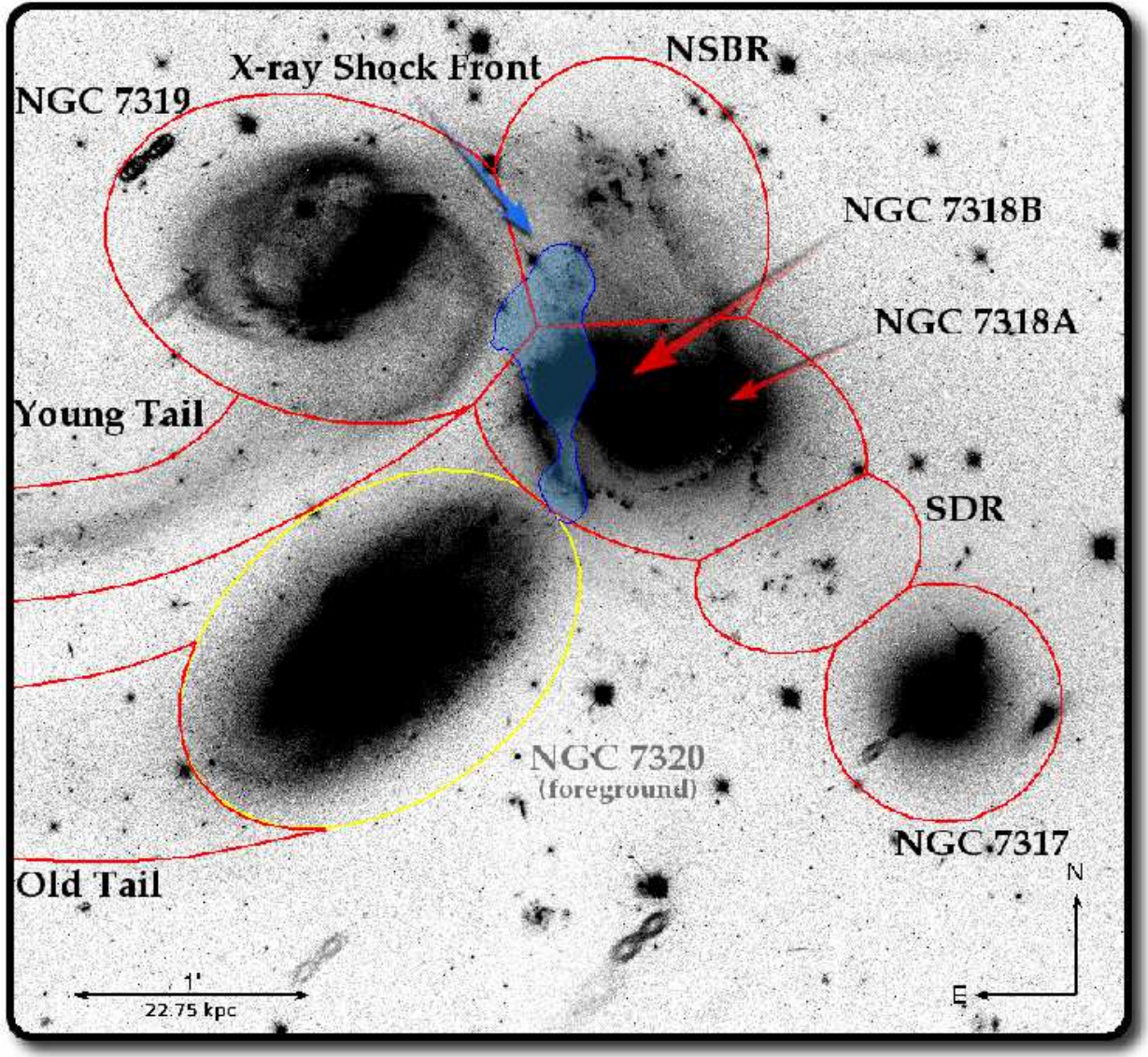}
\end{center}
	\caption{\captionfonts An inverted black and white image (in \vb) with defined regions. The contrast of the image has been scaled to emphasize low surface brightness structures. The 8-shape objects observed in the upper left corner and on the bottom of the image, as well as in Figures~\ref{fig:N7317_SDR} and \ref{fig:N7319}, are ghost images caused by reflections off the CCD and return reflections from the CCD housing entrance window in WFC3.}
\label{fig:finder}
\end{figure}

Stephan's Quintet (hereafter SQ; \altcite{steph1877}), is one of the most studied galaxy groups. There are numerous signs of interactions between the galaxies in SQ, such as galaxy distortions, tidal tails, and active star formation in intergalactic regions, that have happened in the past or are ongoing. However, we still lack a full understanding of the dynamical processes that led to the current spatial distribution of galaxies in SQ. In Figure~\ref{fig:finder}, we present multi-band Hubble Space Telescope imaging of SQ, which covers four galaxies: two spirals (NGC~7319, NGC~7318B) and two ellipticals (NGC~7318A, NGC~7317). In Hickson's designation those galaxies are HCG 92C, -B, -D and -E, respectively. Another spiral galaxy (lower left of the frame), NGC~7320, is known to be a foreground galaxy (\altcite{allen80}, \altcite{moles97}), with a recessional velocity $v_{\rm R}= 739$ \kms~(\altcite{UZC}). We do not examine this galaxy in our study. The core of the group consists of three galaxies NGC~7317, NGC~7319, and NGC~7318A, with approximately the same $v_{\rm R}$ (6637, 6652, and 6671 \kms, respectively; Table~\ref{tab:tab1}). The fourth galaxy, NGC~7318B, is a high-speed intruder (with $v_{\rm R}= 5766$ \kms; Table~\ref{tab:tab1}) that is apparently interacting with the core for the first time (\altcite{mendes94}; \altcite{moles97}). 

The notion that galaxy interactions trigger star cluster formation is the back-bone of this work. Note, however, that although mergers and interactions seem to be necessary for enhanced star formation, they are not a sufficient condition (\altcite{bergvall03}). In the environment of compact groups of galaxies, which combines high densities (comparable to the number densities in the central regions of galaxy clusters) with low velocity dispersions ($\sigma \approx 200$---$300$ \kms), we would expect multiple interactions throughout the history of the group resulting in the production of multiple populations of star clusters. Thus, the history of those interactions can be deduced from detailed studies of the populations of star clusters (\altcite{Gallagher2001} and references within). \citet{Gallagher2001} WFPC2 study, which covered the Young Tail, NGC 7319, and most of NGC 7318A/B system (Figure 1 of their paper), found 115 star cluster candidates. The majority of the candidates were detected not in the central regions of the studied galaxies (NGC~7319, NGC~7318A/B), but in the tidal debris associated with those galaxies and in the Northern Star Burst Region (which we shall describe in greater detail subsequently). They also identified several epochs of recent star formation in SQ, spanning a large range of ages, from the $\sim 2$---$3$ Myr old clusters in the Northern Star Burst Region to the older population of $3$---$12$ Gyr old clusters spread over the entire field of view.

The current study differs from the previous one in its use of a significantly wider field of view, focusing on all of the four related SQ galaxies (including NGC 7317) and their surrounding intragroup medium, and higher sensitivity and spatial resolution, owing to the new Wide Field Camera 3 (WFC3) on the Hubble Space Telescope (\hst). These new observations are approximately 2 mags deeper in the V-band as compared to the previous data of \citet{Gallagher2001}.  The fainter detection limit leads to a larger number of detections, as well as smaller photometric errors.
 
In order to focus our study on a single interaction at a time, we divide the field into eight regions. Three of those regions are named after the galaxies they contain: NGC~7317, NGC~7319, and NGC~7318A/B. The last region contains two galaxies due to their proximity to each other in the plane of the sky. The size of the regions is chosen to include the light from the galaxies as they appear on the \vb\ image where the contrast has been scaled to emphasize low surface brightness structures. Although the gravitational pull of these galaxies stretches beyond the defined borders, we expect this procedure will suffice to identify star cluster candidate populations with their parent galaxies. In their work on the star clusters in nearby starburst galaxies \citet{degrijs03} have required star clusters to have the maximum projected distance of $\sim 1$ kpc from the 3$\sigma_{sky}$ contour in order to still be associated with the given galaxy. If we would apply the same approach in our case, the above mentioned distance of $\sim 1$ kpc from the 3$\sigma_{sky}$ contour will be well within our defined regions for each galaxy. 

Following the naming convention of \citet{Gallagher2001} and \citet{sulentic01}, we also define four regions containing extragalactic features: two tidal tails and two star formation regions. The Old Tail region is located southeast of the center of the group. We can observe only a small part of it emerging from behind NGC~7320. The other tidal tail region, which we call the Young Tail (\S~\ref{sec:YT} ), is located just under NGC~7319, has higher surface brightness, and runs parallel to the Old Tail. If one traces the Young Tail to the East, it points to NGC~7320C, a galaxy out of the WFC3 field of view that may be associated with the group given its accordant velocity, $v_{\rm R}=5985 \kms$, suggesting that this galaxy is responsible for the creation of the tail (\altcite{moles97}). Two tidal arms are found north of NGC~7318A/B. Due to the active star formation in this region, especially in the eastern arm, we call it the Northern Star Burst Region (NSBR; \altcite{Gallagher2001}). It hosts SQ-A (\altcite{xu99}), a strong H$_{\alpha}$-emitting region, which apparently coincides with the overlap region of the two tidal arms. Another star formation region is located between the two galaxies NGC~7318A and NGC~7317. This region, which we call the Southern Debris Region (SDR), is seldom mentioned in the literature as hosting ongoing star formation, although a few studies have shown the presence of H\one\ gas in its vicinity (\altcite{williams02}; \altcite{sulentic01}). The last region, the X-Ray Shock Front, is defined as a contour of soft X-Ray emission ($0.5$---$2.0$ keV) with the value of $0.3$ count pix$^{-1}$ from the {\em Chandra} image in the eastern part of NGC~7318B. This shock wave is the result of a high-speed collision between NGC~7318B, which is blueshifted by $\sim 900$ \kms\ (\altcite{mendes94}) with regards to the average radial group velocity, and the cold intergalactic medium seen in H\one\ (\altcite{shostak84}). The latter was presumably deposited in that location by previous interactions in the group (\altcite{moles97}; \altcite{xu99}). 

Positional, morphological, photometric and redshift information on individual galaxies has been included in Table~\ref{tab:tab1}.

%=============== SECTION 2 ===============
\section{Observations and cluster candidate selection}\label{sec:data}

	\subsection{Data}
The data were obtained with the new WFC3 on \hst, as a part of the Early Release Science (ERS; proposal 11502, PI K.S. Noll). The original observations were carried out in six filters of which we use F438W (\bb), F606W (\vb), F814W (\ib), with the occasional use of the F657N and F658N combined image. The exposure times were $13200$, $5400$, and $7200$ seconds for the \bb, \vb, \ib\ filters, respectively.

	\subsection{Source detection and photometry}
At the distance to SQ of $87.1$ Mpc (adopting $H_{0}=70$ \kms Mpc$^{-1}$), equivalent to a distance modulus of $34.7$ mag, one pixel on the WFC3 corresponds to $\sim 16.5$ pc (0\farcs04 per pixel). Because most individual stars are not luminous enough to be detected at the distance of SQ, we use star clusters as a tracer of star formation. 

In this paper we used PSF fitting as a suitable method of obtaining photometry. This is tested and widely accepted method of getting photometry for unresolved star clusters (\eg\ \altcite{gall10}, \altcite{konst10}) and is very useful in the varying background and crowded field conditions present in SQ.

According to the WFC3 Instrument Handbook the average FWHM of the PSF for BVI filters is 1.7 pixels. After image manipulations the size of PSF only gets larger. Thus, assuming the best case scenario of FWHM being 1.7 pixels, in order for a star cluster to be larger than the PSF it should have physical size of $\sim 27$ pc whereas the average size of a star cluster is currently accepted as 4 pc (\eg, \altcite{barmby06}; \altcite{scheep07}).  Thus, the star clusters of SQ are expected to be unresolved and should appear as point sources, hence the detection and careful selection of point sources are essential to our study.

Point sources were detected with the \texttt{DAOFIND} (\altcite{stetson87}) task in \texttt{IRAF}\footnote{IRAF is distributed by the National Optical Astronomy Observatory, which is operated by the Association of Universities for Research in Astronomy (AURA) under cooperative agreement with the National Science Foundation.}, with a $0.6\sigma$ threshold on a median-divided image (we used a 13$\times$13 pixel smoothing window and divided the original image by the smoothed one). The detection process closely follows that described in \citet{gall10}. 

The PSF was constructed from bright, isolated and unsaturated stars with point-like radial curves of growth. There were 27, 14, and 48 of such stars in \bb, \vb,\ and \ib\ filters, respectively. This procedure was repeated for each filter. Using the \texttt{DAOFIND.ALLSTAR} package in \texttt{IRAF} the photometry for all point sources was obtained. The aperture corrections between 3 to 10 pixels were calculated for all filters, as an average of the difference between magnitudes obtained from aperture photometry with a 10-pixel aperture and PSF-magnitudes (calculated at 3 pixels) for all the PSF stars ($\Delta B_{3\rightarrow 10} = 0.207$ mag, $\Delta V_{3\rightarrow 10}=0.258$ mag, $\Delta I_{3\rightarrow 10} = 0.342$ mag). Using the on-orbit enclosed energy (EE) curves for WFC3 UVIS we calculated the 10-pixels to infinity correction as the difference between unity and the enclosed energy in the given aperture and wavelength ($\Delta B_{10\rightarrow \infty}=0.110$ mag, $\Delta V_{10\rightarrow \infty}=0.103$ mag, $\Delta I_{10\rightarrow \infty}=0.108$ mag). The photometry in each filter was corrected for foreground extinction $E(B-V)=0.079$ mag (\altcite{schlegel98}), with values of $A_{438}=0.32$ mag, $A_{606}=0.22$ mag, and $A_{814}=0.14$ mag. For the final catalog, we require sources to be detected in all three broad-band filters, which eliminates most contamination from spurious detections.

	\subsection{Star Cluster Candidate Selection}\label{sec:SCCs}
In order to select star cluster candidates (SCC), we require sources to pass several additional criteria.  First, we reject sources with colors \bb\ $-$ \vb\ $>1.5$ mag and \vb\ $-$ \ib\ $>1.0$ mag, in order to eliminate most foreground stars in the Galaxy. In order to select point sources, we follow the prescription described in \citet{rejkuba05}, selecting sources which have: (1) magnitude error $\sigma_{mag} \leq 0.3$ mag in all three bands, (2) the sharpness of the source (a measure of the relative width of a source with respect to that of the PSF) between $-0.2$ and $0.2$ in all three bands, and (3) the goodness of fit factor from PSF-fitting $\chi \leq 3$ in \ib\ band. The use of \ib\ band for the $\chi$ filter is dictated by the fact that the PSF model is best determined in that band and there is no expected contamination from emission lines (\S \ref{sec:contam}). Finally, we only consider sources brighter than $M_{V_{606}}$ of $-9$ mag, which should eliminate nearly all individual luminous stars in SQ (\altcite{whitmore99}; 2010).

Based on these criteria, we select $64$ sources in NGC~7317 region, $89$ in NGC~7319, $133$ in NGC~7318A/B, $41$ in the regions of the X-ray Shock Front, $25$ in the Young Tail, $33$ in the Old Tail, $110$ in the Northern Star Burst Region (NSBR), and $42$ SCCs in NGC~7317 (see Table \ref{tab:res}).  

	\subsection{Completeness}\label{sec:complete}
In order to test the completeness of our final list of SCCs, we used \texttt{DAOPHOT.ADDSTAR} to add 5000 artificial point sources to the image in the magnitude range of $24$---$28$ mag, i.e. $-10.7$ mag to $-6.7$ mag in absolute magnitude at the distance of SQ. We then followed the detection algorithm outlined in section~\ref{sec:SCCs}. We found that for recovery rates of $50$\%\ and $90$\%\ the limiting magnitudes were $27.6$ mag and $26.5$ mag in the \bb\ band, $27.7$ mag and $26.6$ mag in the \vb\ band, and $27.5$ mag and $26.5$ mag in the \ib\ band (after aperture and extinction corrections were applied).

Though the quoted completeness levels are for the entire field, the star clusters are clearly concentrated within the galaxies, where the background can be high and variable.  Nevertheless, the completeness levels within the regions defined in Figure~\ref{fig:finder} were $\ge 90$\% for a star cluster with $M_{\rm V}=-9$, the cutoff for inclusion in our catalog.

	\subsection{Contamination}\label{sec:contam}
\begin{figure}
\vspace{-0.75cm} 
\begin{center}
	\includegraphics[width=1.0\textwidth]{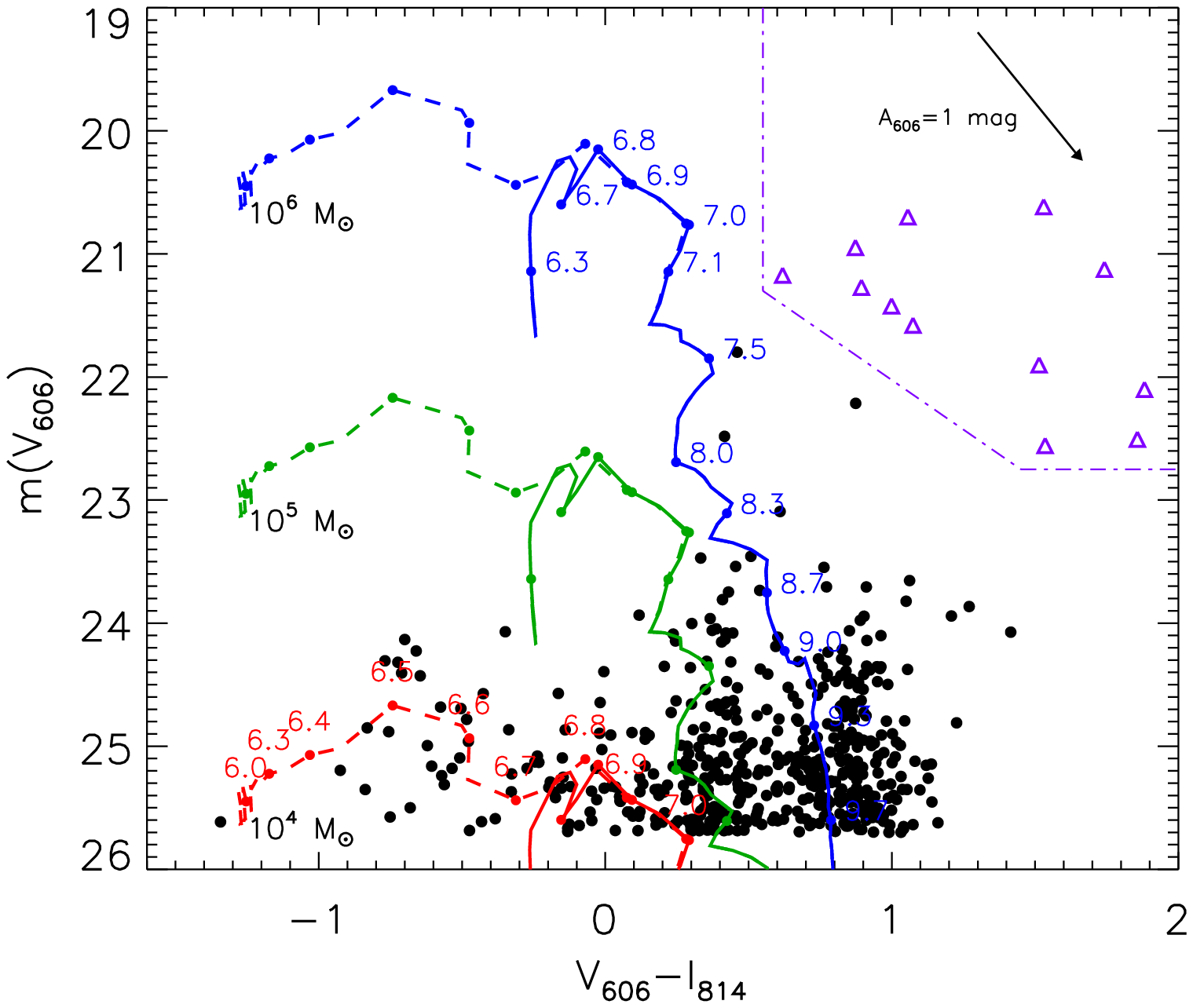}
\end{center}
    \caption{\captionfonts \vb\ versus \vb\ $-$ \ib\ color-magnitude diagram for SCCs (black filled circles) in all regions. The solid lines are evolutionary tracks from \citet{marigo08} with a metallicity of $\frac{1}{5}$~\Zsun (\altcite{sqmetal}) for different cluster masses (in Solar masses). The dashed lines represent evolutionary tracks that account for nebular emission (SB99), as suited to the very youngest clusters that have not yet expelled their natal gas (see \S \ref{sec:evoltrack}). The objects represented by purple triangles were spectroscopically confirmed to not be members of SQ (\S \ref{sec:contam}). We cannot identify this contamination with the color-color plot (Figure \ref{fig:cc}) since most of the contaminating objects are mixed with our SCCs. However, from a color-magnitude diagram we can see that the vast majority of these contaminating objects would only affect the old massive cluster bin. A Galactic extinction vector of length 1~mag in $V_{606}$\ is shown in the top right corner.}
\label{fig:cmd}
\end{figure}

\begin{figure}
\vspace{-0.75cm} 
\begin{flushleft}
	\includegraphics[width=1.0\textwidth]{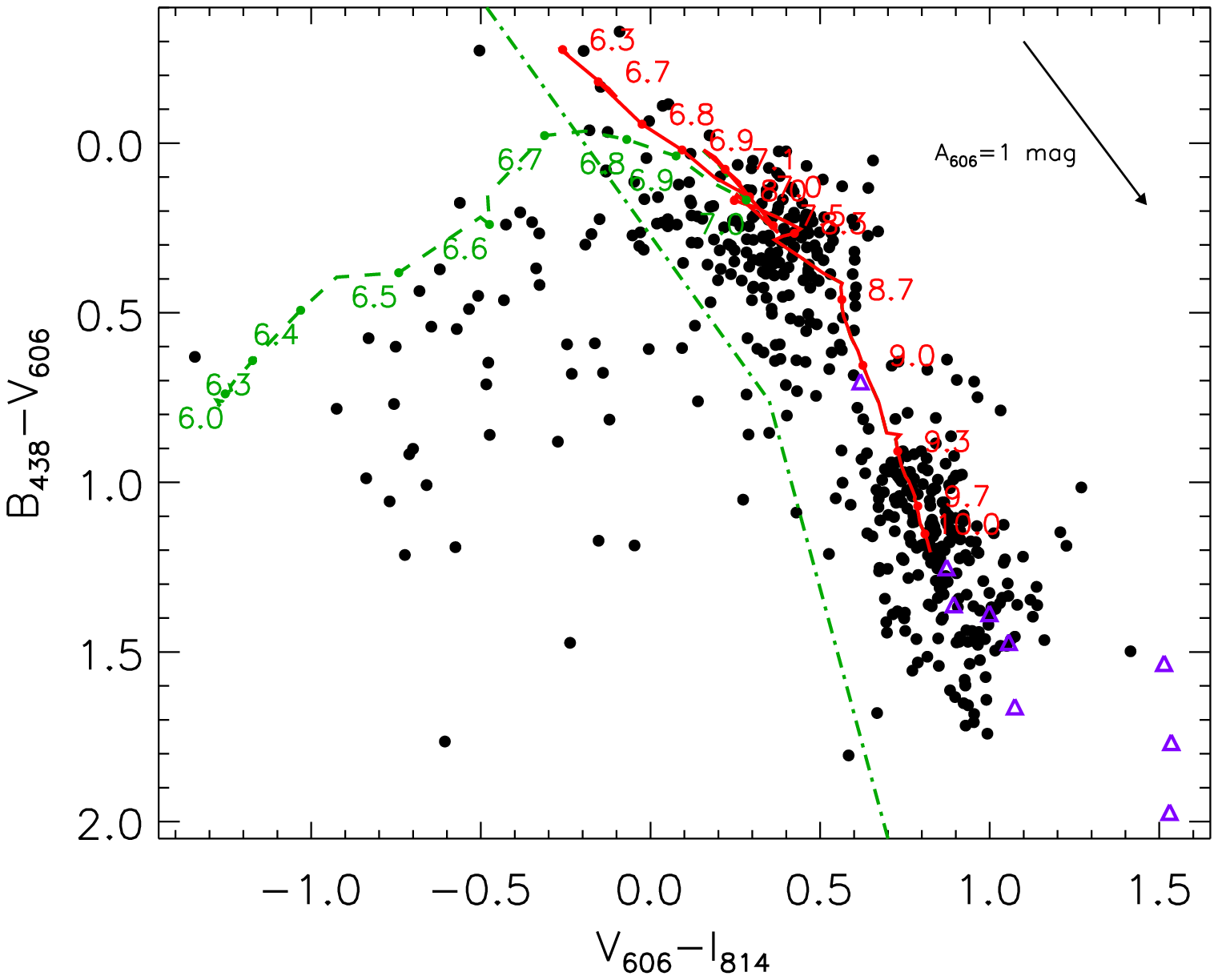}
\end{flushleft}
    \caption{\captionfonts \bb\ $-$ \vb\ versus \vb\ $-$ \ib\ colors for SCCs in all regions, plotted on top of an evolutionary track (solid red line; \altcite{marigo08}) with a metallicity of $\frac{1}{5}$~\Zsun (\altcite{sqmetal}). SCCs are denoted by filled black circles. The objects represented by purple triangles were spectroscopically confirmed to not be members of SQ (\S \ref{sec:contam}). The dashed green line shows an evolutionary track that accounts for nebular emission (SB99). All data-points that lie to the left of the dot-dashed green line are consistent with the nebular tracks, accounting for reddening along the Galactic extinction vector shown in the top right. The dot-dashed green line is drawn based on the width of the distribution of SCCs with respect to the evolutionary track and taking into account typical photometric errors.}
\label{fig:cc}
\end{figure}

From a sister spectroscopic study to this that is being conducted in parallel (G. Trancho et al., in prep.), we obtained a list of objects that were found not to be associated with SQ, such as foreground stars and background quasars. The identifications of the objects were determined with spectra from GMOS on Gemini North (programs GN-2004B-DD-8 and GN-2006A-Q-38), based on the measured $v_{\rm R}$. As can be seen from the $M_{V_{606}}$ vs. \vb\ $-$ \ib\ color-magnitude diagram of Figure~\ref{fig:cmd}, the contaminating objects have \vb\ $-$ \ib\ colors redder than $0.6$ mag. Although the spectroscopic sample of the study is luminosity-limited, the vast majority of contaminating objects would only affect the old globular cluster bin (Figure~\ref{fig:cc}). Also, the spatial distribution of the SCCs in our sample coincides with the host galaxies, and features in the intragroup medium (such as tidal arms and star burst regions), as we would expect from legitimate star clusters. Additionally, we estimated the contamination from the Galactic foreground stars based on the results from the Besan\c{c}on models\footnote{http://model.obs-besancon.fr/} (\altcite{robin03}) in the direction of SQ up to the distance of 100 kpc. The model predicts $\sim 80$ foreground stars in the field of view of SQ ($0.007$ deg$^2$) within the magnitude and color ranges of SCCs. Hence, for the area of NGC~7319 region, which has the largest area of all the regions, we expect to find about 5 foreground stars with apparent magnitudes ranging from $19$ to $29$ in all three bands.

Taking into consideration all of the above, we are confident that our final list of star cluster candidates is not heavily contaminated by foreground/background objects.

	\subsection{Evolutionary Tracks}\label{sec:evoltrack}
We compare the photometry of our SCCs with predictions from stellar synthesis models from \citet{marigo08}, assuming the \citet{kroupa98} stellar initial mass function, with upper and lower masses of 0.15 and 100 \Msun, and $\frac{1}{5}$ \Zsun, as measured by \citet{sqmetal} with their X-ray study. However, these models only include stellar contributions, not contributions from nebular emission lines. Nebular emission can be strong in regions with active star formation, and can strongly affect measurements in the \vb\ filter (with H$_{\alpha}$, H$_{\beta}$ and [O\three] emission lines) and less so in the \bb\ filter (with the [O\three] and H$_{\beta}$ emission lines). Because of these contributions, the \bb\ $-$ \vb\ colors appear redder and the \vb\ $-$ \ib\ colors appear bluer (\eg~\altcite{vacca92}, \altcite{conti96}) than the colors from continuum emission alone. In order to account for nebular emission from the youngest star clusters, we include this contribution by calculating the expected strength of the H$_{\alpha}$ and H$_{\beta}$  emission lines from \texttt{Starburst99} (\altcite{leith99}) for the same IMF and metallicity. The [O\three] line strength is calculated as $0.7\times$ H$_{\beta}$, which is the largest ratio observed for the ratio of [O\three] to H$_{\alpha}$ in the KISS sample of low-mass star forming galaxies (\altcite{salzer05}).

%=============== SECTION 3 ===============
\section{Results and Discussion}\label{sec:resdis}
\begin{figure*}
\begin{center}
	\includegraphics[width=0.78\textwidth]{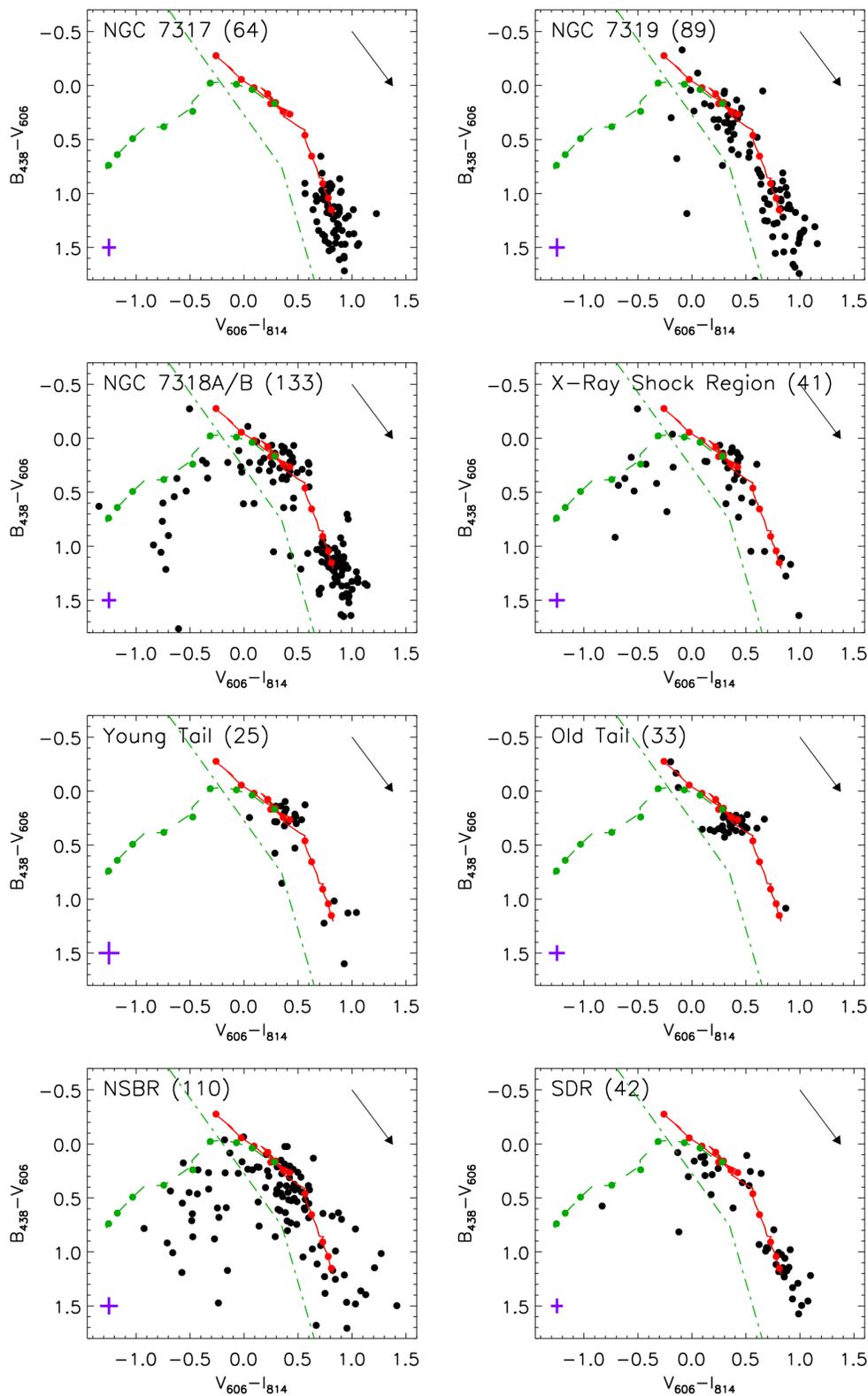}
\end{center}	
    \vspace{-0.35cm} 
	\caption{\captionfonts Color-color plots for Star Cluster Candidates (SCCs) in individual regions, as defined in Figure~\ref{fig:finder}. The number in parentheses gives detected SCCs for that particular region. A typical photometric error bar, based on the median errors, is located in the lower left corner of each panel. The fainter the population, the larger the typical errors. Larger versions of the color-color diagrams are presented in Figures~\ref{fig:N7317_SDR} --- \ref{fig:NSBR}, along with the spatial distribution of the SSCs.}
\label{fig:colours}
\end{figure*}

\begin{figure}
\begin{center}
    \subfigure{\includegraphics[width=0.90\textwidth]{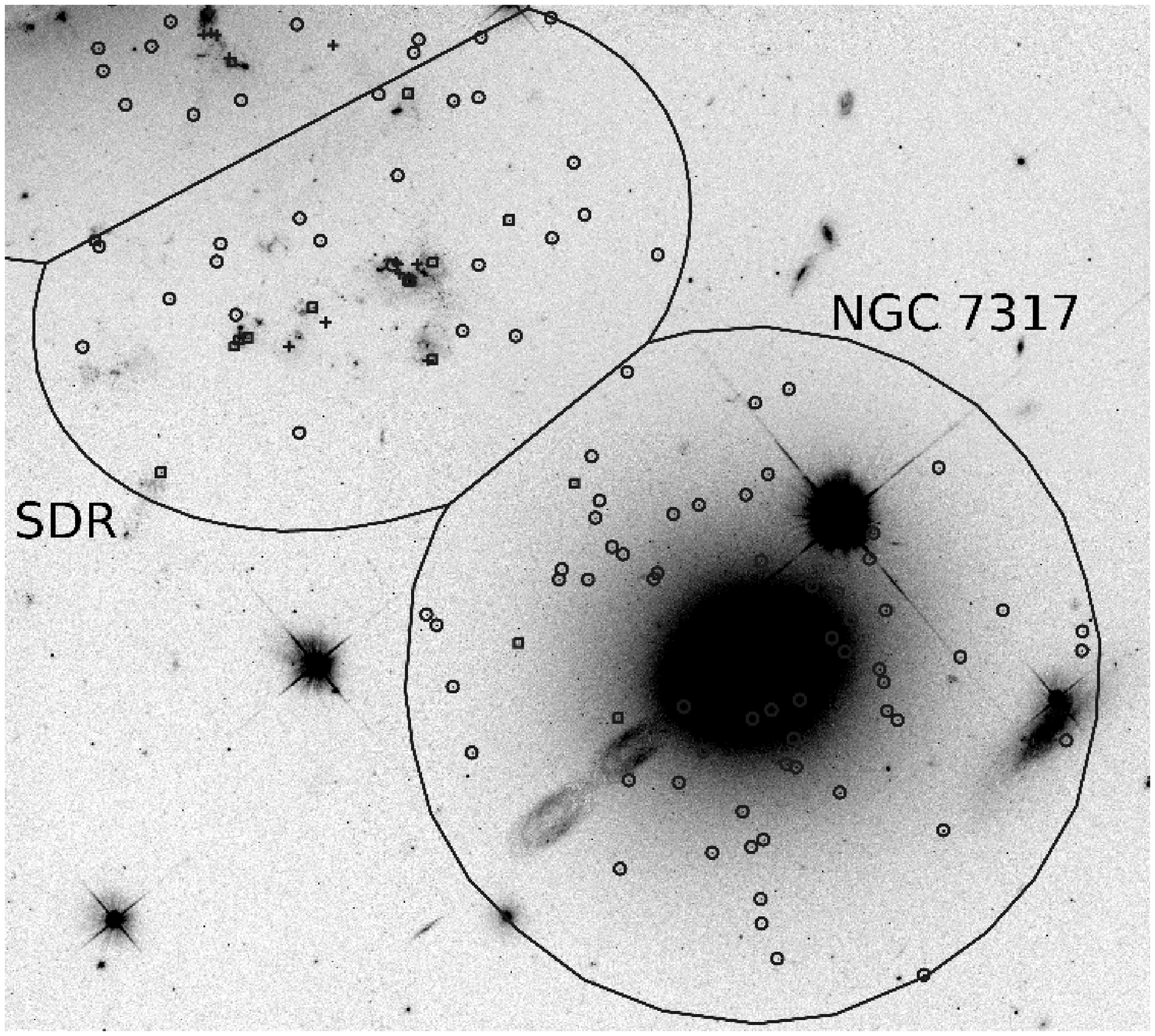}}\\
    \subfigure{\includegraphics[scale=0.58]{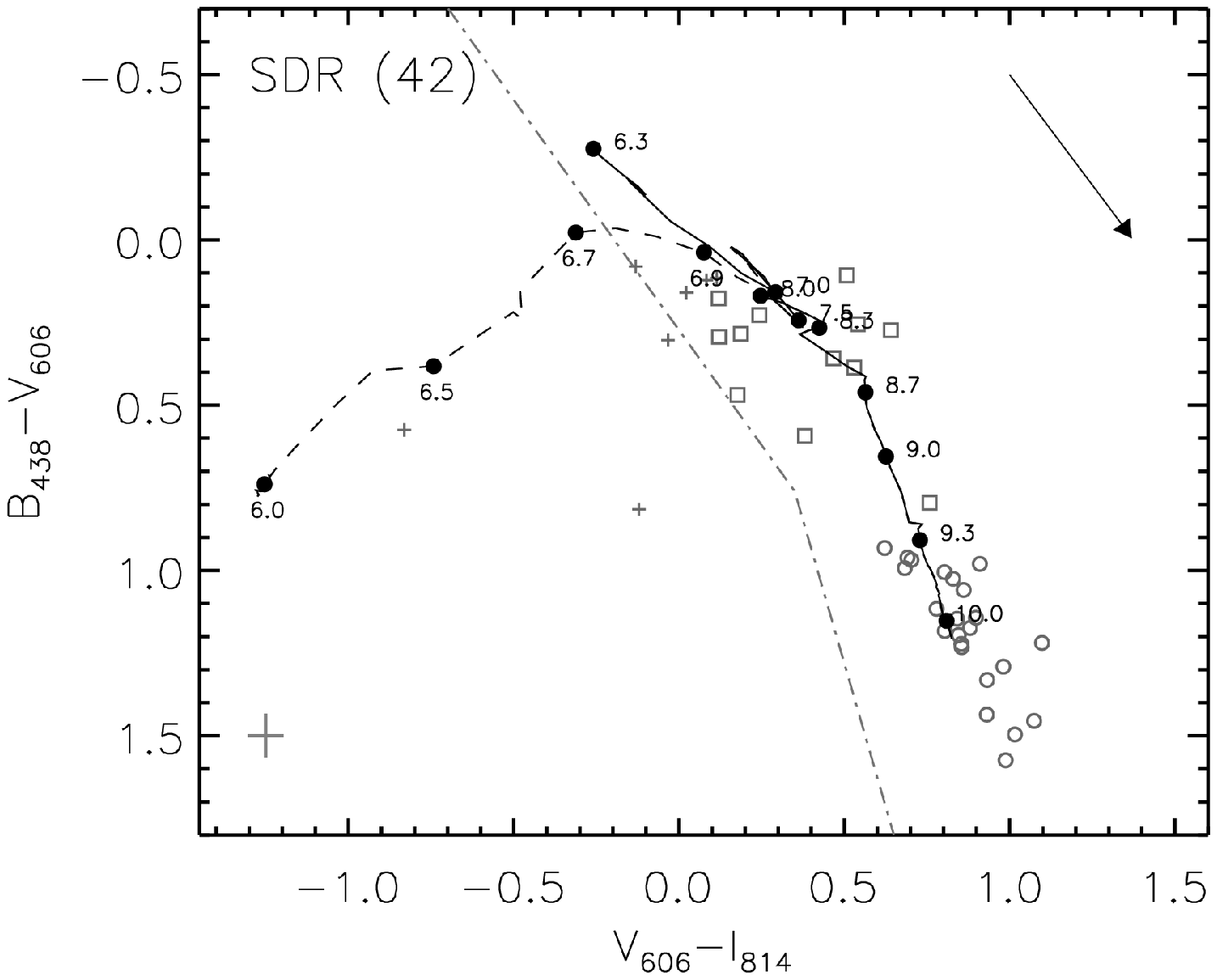}
    		   \includegraphics[scale=0.58]{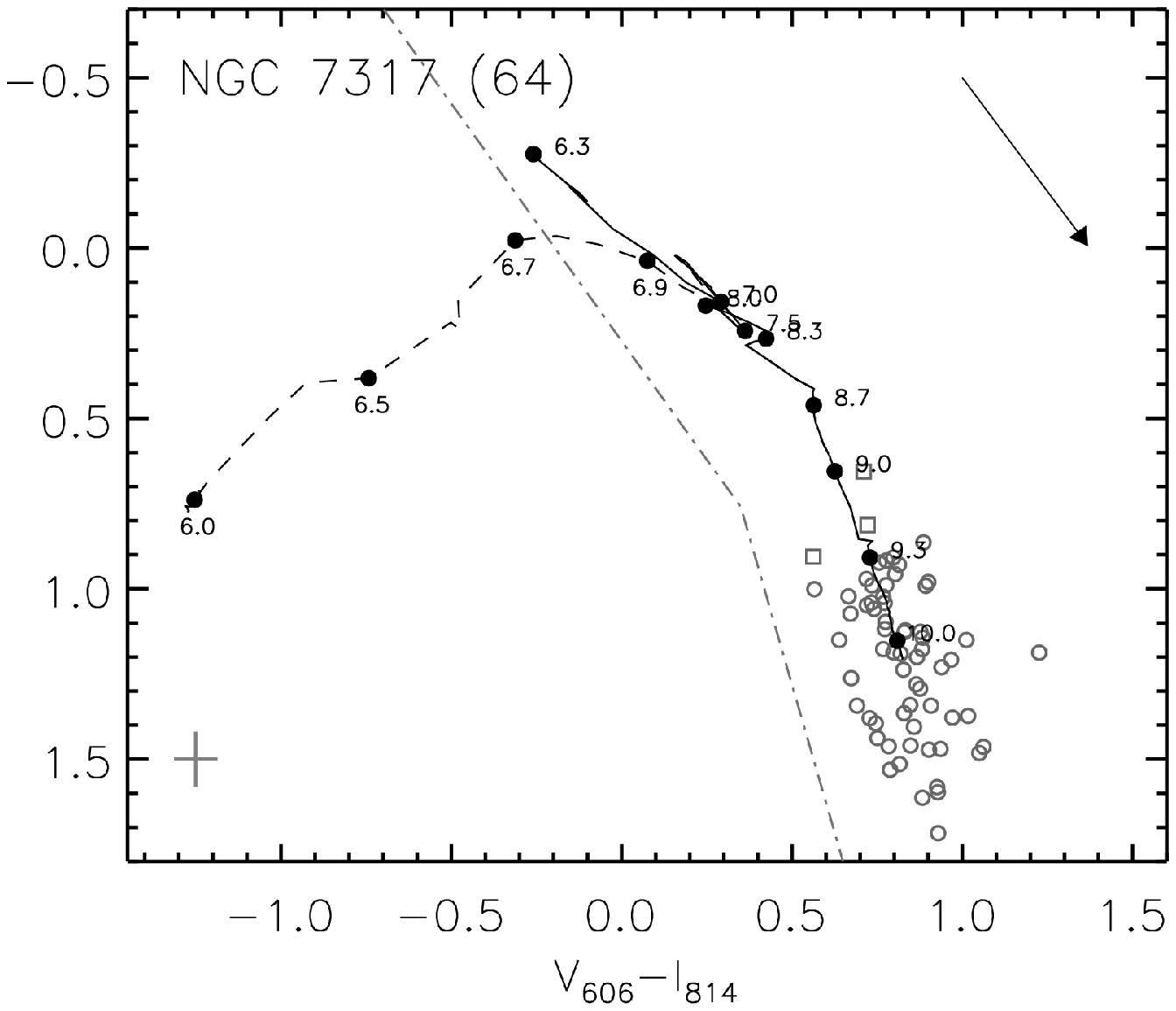}}
\end{center}  
    \vspace{-0.75cm} 
    \caption{\captionfonts {\bf Top}: The spatial distribution of SCCs in the SDR and NGC~7317 regions marked according to their ages: crosses represent very young star clusters ($\lesssim 10$ Myr), squares represent star clusters from $\sim 10$---$20$ Myr to 2 Gyr old, and circles represent star clusters that are $\gtrsim 2$ Gyr old. {\bf Bottom}: Color-color plots of the same regions. In parentheses are the numbers of detected SCCs in that regions. SCCs in the SDR (bottom left panel) are divided into two groups, the old star clusters and cluster candidates with ages between $50$ to $200$ Myr, with virtually no SCCs between those two groups. NGC~7317 (bottom right panel) hosts mostly old star clusters as expected for an elliptical galaxy.}
\label{fig:N7317_SDR}
\end{figure}

Below we discuss the results obtained for each region of SQ, defined earlier in \S \ref{sec:intro} and \mbox{Figure \ref{fig:finder}}. Most of our conclusions come from analysis of the color-color plots for those regions. Figure \ref{fig:cc} presents such a plot for all the SCCs ($M_{V_{606}} < -9$ mag) detected in SQ, and Figure \ref{fig:colours} presents color-color plots for each of the defined regions, so that the reader can compare them side-by-side. To better understand the spatial distribution of star clusters with regard to their approximate ages we also included zoomed-in images of the various regions in \vb\ band and their associated color-color plots in Figures \ref{fig:N7317_SDR} - \ref{fig:NSBR}.

Intrinsic reddening will mimic older cluster ages in broad-band colors, i.e. it will move the broad-band colors redward along the model predictions. Since we do not have an extinction map for SQ we will not include the effects of extinction explicitly. Rather we will be making qualitative statements based on the visual spatial distribution of the dust lanes. However, to indicate possible effects of intrinsic reddening we have included an A$_{V_{606}} = 1$ mag Galactic reddening vector in all color-color and color-magnitude diagrams. This might not be appropriate for the SQ galaxies considering their lower metallicities (\altcite{sqmetal}), but is conservative and is widely adopted in the literature. The derived ages therefore can be considered as upper limits; if local extinction is small the obtained ages are closer to the real ones, and if local extinction is large then obtained ages are probably higher than they should be. The broken dot-dashed green line present on the color-color plots is drawn to roughly parallel the evolutionary tracks at a distance set to respect the width of the observed distribution of SCCs along the tracks.  The spread in the distribution includes both intrinsic and photometric scatter.  The SCCs that are located to the left of the broken dot-dashed green line are likely to be very young (less than $10$ Myr), due to nebular emission that affects their colors.

For each of the regions we summarize the available literature trying to combine the theoretical models and computer simulations, especially \citet{renaud10}, the most recent and detailed dynamical model to date, with our observations to construct a comprehensive history of interactions in SQ during the last Gyr.

	\subsection{NGC~7317}
NGC~7317 is an elliptical galaxy and, as expected for a galaxy of that type, hosts mostly old star clusters. The distribution of these clusters in color space is concentrated near \bb\ $-$ \vb\ $= 1.1$ mag and \vb\ $-$ \ib\ $= 0.8$ mag (Figure \ref{fig:N7317_SDR}). The minimal reddening  that would be expected in NGC~7317, as well as the spatial distribution of the clusters, which are concentrated around the core of the galaxy, suggest that a large number of them can be considered globular clusters. 

	\subsection{Old Tail}\label{sec:OT}
The Old Tail is a low surface brightness tidal structure that originates presumably from NGC~7319 (\altcite{renaud10}) and formed through the interaction with NGC~7320C (\altcite{moles97}; \altcite{sulentic01}). The Old Tail is mostly obscured by NGC 7320 in our field of view, and as such  it is heavily contaminated by the objects of the foreground galaxy NGC~7320. However, a small unobstructed part of that tidal feature, which we are able to observe just southeast of NGC~7320, appears to have minimal contamination from  NGC~7320. 

\begin{figure}
\begin{center}
    \subfigure{\includegraphics[width=0.825\textwidth]{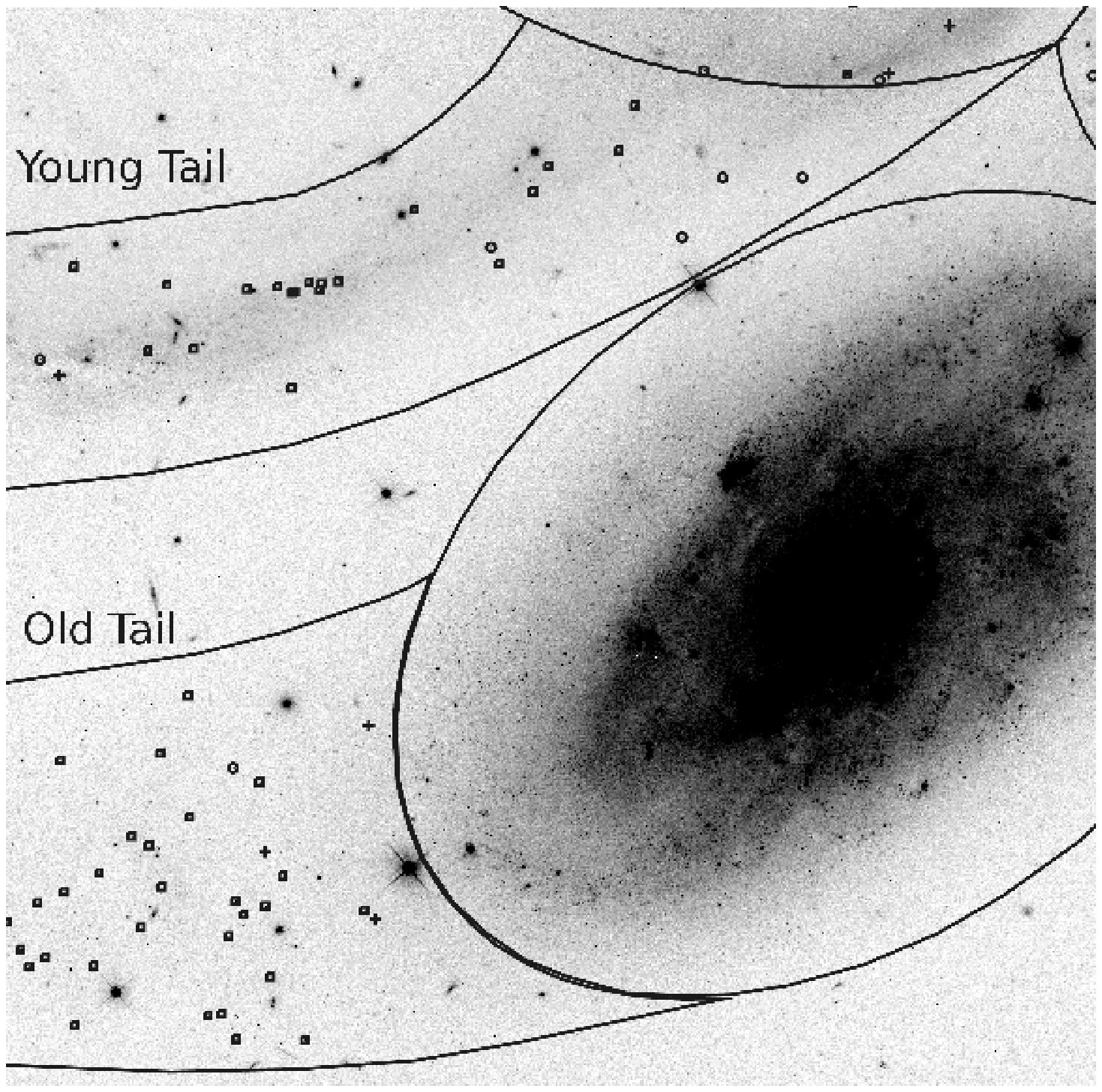}}\\
    \subfigure{\includegraphics[scale=0.58]{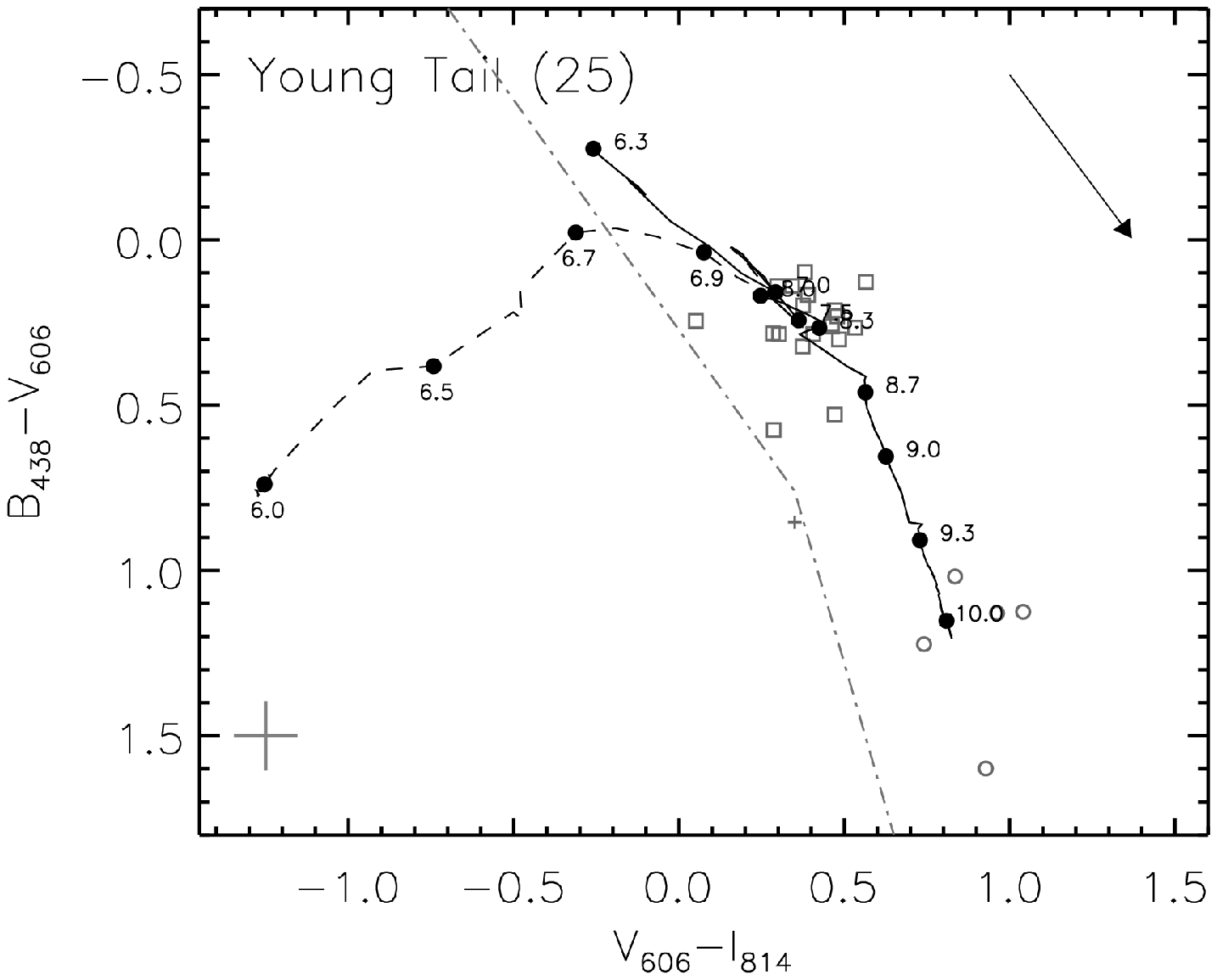}
    		   \includegraphics[scale=0.58]{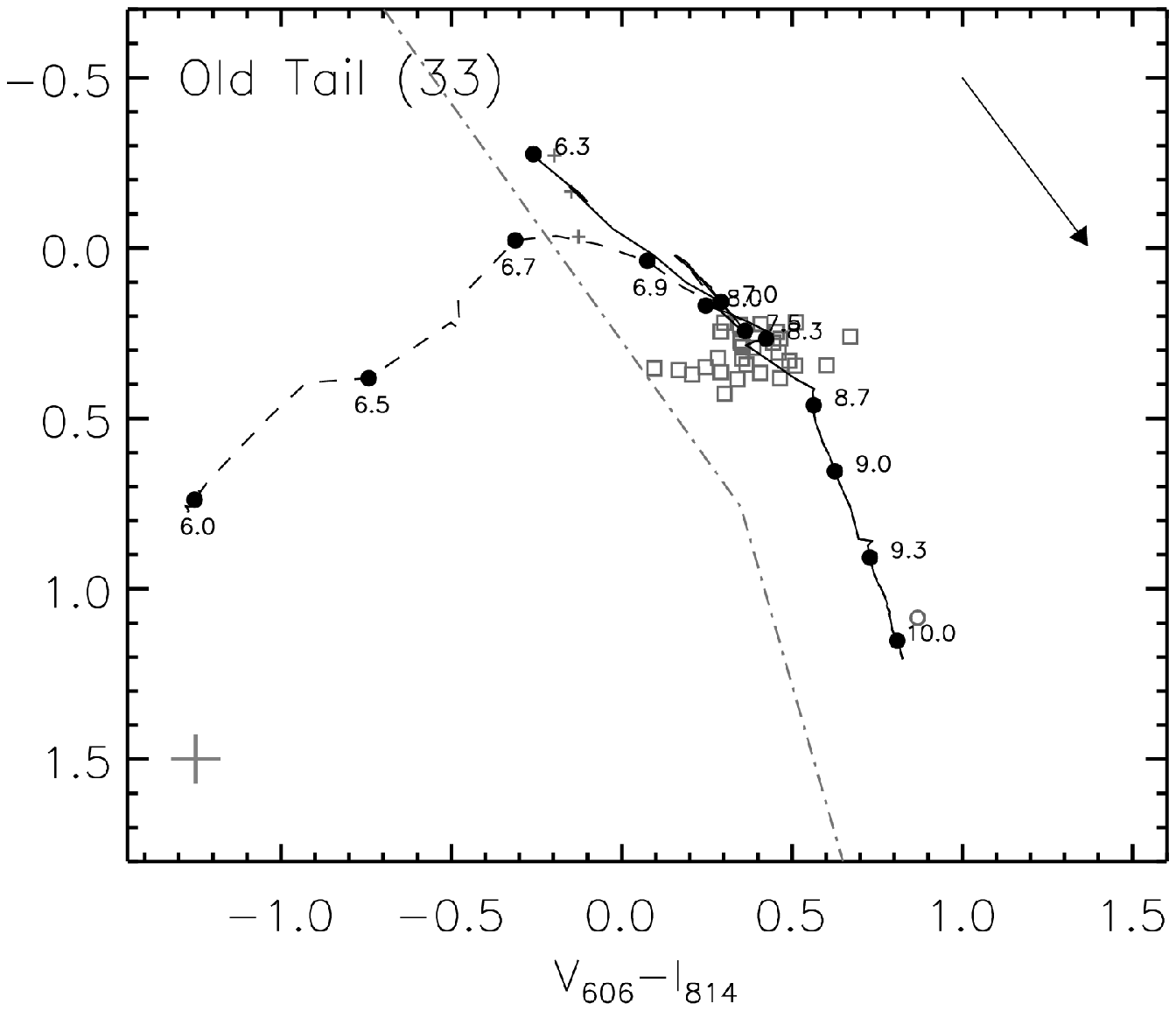}}
\end{center}
  \vspace{-0.75cm} 
  \caption{\captionfonts {\bf Top}: The spatial distribution of SCCs in the Young and Old Tail regions marked according to their ages (symbol definitions follow Figure \ref{fig:N7317_SDR}). {\bf Bottom}: Color-color plots of the same regions. For the Young Tail (bottom left panel) the concentration of blue clusters with mean colors of \bb\ $-$ \vb\ $= 0.22 \pm 0.07$ mag and \vb\ $-$ \ib\ $= 0.39 \pm 0.12$ mag is observed. These mean colors most closely match a model age of $200$ Myr. Clusters in the Old Tail (bottom right panel) have a narrow range of colors, with a mean \bb\ $-$ \vb\ $= 0.31 \pm 0.06$ mag and a mean \vb\ $-$ \ib\ $= 0.38 \pm 0.12$ mag, corresponding to a model age of $400$ Myr.}
\label{fig:YT_OT}
\end{figure}

Indeed, if star clusters of the Old Tail were part of the NGC~7320 then there should be some explanation to such a high, asymmetric concentration of SCCs outside the galaxy (Figure~\ref{fig:YT_OT}). The most plausible explanation would involve an interaction of NGC~7320 with another galaxy. However, the H\one\ study of SQ by \citet{Gutierrez2002} did not find any evidence (tidal tails, bridges, etc.) of ongoing or past interactions between NGC~7320 and NGC~7331, the closest galaxy (projected separation is $\sim 100$ kpc) with similar recession velocity $v_{\rm R} \sim 800$ \kms. Moreover, the observations of H\one\ in NGC~7320 conducted by \citet{williams02} revealed the signature of normal disk rotation, without any signs of tidal distortion.

To further test the membership of SCCs in the Old Tail region we looked at their  size distribution. We used \texttt{ISHAPE} software (\altcite{larsen99}) to determine the FWHM for all detected SCCs in the Old Tail and NGC~7320 with an observed magnitude range of \vb\ $ = 23.5$ --- $25.6$ mag. The distribution of radii for the Old Tail sources is quite narrow (from $1.0$ --- $2.2$ pix), and is peaked at r $\sim 1.7$ pix.  In contrast, the NGC 7320 sources have a much broader distribution of radii ranging from $0.2$ --- $4.4$ pix with a peak at r $\sim 1.9$ pix. Given that the distance modulus of NGC~7320 is $29.1$ mag, the distribution likely contains a mix of stars and resolved clusters. A Kolmogorov Smirnov test applied to the distributions of FWHM gave $< 0.004$ probability that both sets of SCCs came from the same distribution.  

All of the above suggests that the SCCs in question do belong to the Old Tail and contamination from NGC~7320 is minimal.  Also, clusters in the Old Tail have a narrow range of colors (the color and spatial distributions are shown in Figure~\ref{fig:YT_OT}), with a mean \bb\ $-$ \vb\ of $0.31\pm0.06$ mag (where the error is the standard deviation in the mean) and a mean \vb\ $-$ \ib\ of $0.38\pm0.12$ mag.  This is another sign of minimal contamination from the foreground galaxy, since variable extinction that is expected in spiral galaxies would tend to spread out the colors along the extinction vector.  The uniform and relatively young age distribution is also suggesting in situ star formation, meaning that the Old Tail was formed with a large amount of gas.  Because the tidal tail environment is usually associated with little extinction, the predicted age that best matches the mean colors is $400$ Myr.  This is consistent with ages obtained by \citet{xu05}: $t\simeq10^{8.5\pm0.4}$ yr, corresponding to the age range of $125$---$800$ Myr, from their extinction-corrected UV and optical colors, although somewhat less consistent with ages $\gtrsim 500$ Myr, the estimate based on dynamical arguments (\altcite{xu05}).

We also tried to detect the Old Tail SCCs situated behind the foreground galaxy NGC~7320. Although we found a number of sources with a consistent spatial distribution and with similar colors to our uncontaminated Old Tail region, no definite results could be drawn because the projected width of NGC~7320 is almost identical to the apparent width of the Old Tail and contamination by star clusters or stars with the same colors in the foreground galaxy cannot be reliably determined.

	\subsection{Young Tail}\label{sec:YT}
The Young Tail is a tidal structure that is parallel to the Old Tail but has higher surface brightness. Clusters in the Young Tail tend to have bluer colors than those in the Old Tail, although there are a few redder SCCs as well. The concentration of blue clusters (Figure~\ref{fig:YT_OT}) have mean colors of \bb\ $-$ \vb\ $= 0.22\pm0.07$ mag and \vb\ $-$ \ib\ $= 0.39\pm0.12$ mag. These mean colors most closely match a model age of $200$ Myr, indicating that the bulk of the clusters formed more recently than in the Old Tail. This age estimate is in good agreement with the values of $\sim 150$ Myr, obtained by \citet{Gallagher2001} from their WFPC2 imaging study. Also, the sister spectroscopy study led by G. Trancho (in preparation) constrained the ages of clusters in the Young Tail to $\lesssim 200$ Myr. The redder, and hence presumably older clusters (3 clusters with ages $\sim 500$ Myr and 5 clusters $\gtrsim 2$ Gyr old), present in the Young Tail were plausibly deposited there from the disks or halos of the galaxies by the interaction responsible for the creation of this tidal feature; alternatively, they are a line-of-sight projection of a few old clusters that surround the nearby galaxies. Another possibility is that these clusters are young and highly extincted, although this explanation is less likely as tidal tails typically have low extinction.

Previous studies (\altcite{moles97}; \altcite{sulentic01}) credit a second pass of NGC~7320C through NGC~7319 as the event that formed the Young Tail. However, there are a few problems with this hypothesis. Figure 5 in \citet{williams02} presents the total H\one\ column density distribution in SQ. Based on that figure, the distribution of H\one\ in the Young Tail is not continuous. Approximately two thirds of the optical tail that is closer to NGC~7319 is H\one\ free; the last third contains H\one\ and the density contours of the gas follow the optical tail. However, at the end of the optical tail the H\one\ contours change direction very rapidly and align themselves along the North-South line. The authors point out a common tendency that is seen in numerical studies of galaxy interactions: tidal tails are usually pointing to the source of disturbance. This happens due to the exchange of momentum between the intruder-galaxy and the tidal debris. From this observation and the H\one\ distribution, they conclude that the H\one\ associated with the last third of the optical part of the Young Tail cannot be primarily driven by NGC~7320C which lies to the East.

In \citet{sulentic01}, the observed lack of H\one\ in the region of the tail closest to NGC~7319 and the length of the feature are explained by a low-velocity and low inclination interaction between NGC~7319 and NGC~7320C. However, \citet{renaud10} argue that in order to strip such a large amount of material, both gaseous and stellar, the low mass NGC~7320C had to pass very close to the center of NGC~7319 where the density is high. Furthermore, they argue that because of dynamical friction, orbital decay after two encounters would be too large to allow NGC~7320C to occupy its current, distant position. In order to verify this hypothesis, they conducted 96 simulations of double-encounters between NGC~7319 and NGC~7320C, varying the orbital parameters within reasonable values, and no model gave the observed separation between these galaxies.

A similar point is argued by \citet{xu05}. Based on the measured redshifts of NGC~7319 and NGC~7320C they estimated that it would take more than $500$ Myr for NCG~7320C to get to its current position after the encounter with NGC~7319. This is consistent with the Old Tail age but is too old for the Young Tail. Instead, they propose the scenario where the Young Tail is formed by a close encounter between NGC~7319 and NGC~7318A.

Another possible argument against NGC~7320C being responsible for the formation of the Young Tail is the appearance of that galaxy. Assuming this is the second interaction between these galaxies, and taking into account the difference between the masses of these two galaxies (NGC~7319 is $\sim 30$---$40$ times more massive; \altcite{renaud10}), and the disturbed appearance of NGC~7319, we would expect NCG 7320C to look at least as disturbed as NGC~7319 or, in all probability, to be severely tidally distorted. Although NGC~7320C does exhibit signs of past interactions (\eg, strong spiral arms) they are not as strong as in the case of NGC~7319. This, however, could be an effect of particular orbit/spin orientations. The models of the encounter between the two galaxies (mass ratio 4:1; not 30:1 as in the case of NGC~7319 and NGC~7320C) showed that, in the case of prograde (for the larger galaxy) and retrograde (for the smaller galaxy) encounter, the larger galaxy can develop a tidal feature while the smaller one would be left seemingly unaffected (Renaud, 2010, private communication).

Based on the reasons mentioned above, we agree with the conclusion of \citet{renaud10}, from collisionless N-body simulations of SQ, that the Young Tail was formed by the interaction between NGC~7319 and NGC~7318A, and we further advance their scenario with a more robust age restriction for this event from the SCC population to $150$---$200$ Myr ago.  

	\subsection{NGC~7318A/B}\label{sec:AB}
Due to the spatial proximity of NGC~7318A and B, we have decided to analyze them as a unit rather than as individual galaxies. 

From the color-color diagram in Figure~\ref{fig:N7318AB}, we conclude that the NGC~7318A/B region exhibits ongoing star formation for the last $250$ Myr. In the literature NGC~7318A is identified as an elliptical galaxy  (\eg~\altcite{nilson73}; \altcite{rc3}; \altcite{moles98}), which would explain the relatively large number of old SCCs observed in this region. Another interesting feature present in the color-color plot for NGC~7318A/B (Figure~\ref{fig:N7318AB}) is a lack of detected SCCs between the ages of $400$ Myr and $2$ Gyr. Five out of the seven objects that are located in the box of $0.0<$ \vb\ $-$ \ib\ $<1.1$ and $0.5<$ \bb\ $-$ \vb\ $<0.8$ (i.e., approximately 1 Gyr old) are either coincident with bright sources in the $H_{\alpha}$ images or located in the actively star-forming regions, making them very likely to be reddened young clusters. Thus, there appears to be a well-pronounced gap in color space between the ages of $\sim 400$ Myr and $2$ Gyr. Moreover, the presence of this gap is somewhat unusual as recent studies (\eg\ \altcite{gall10}, \altcite{konst10}) have shown that the distribution of SCCs for spiral galaxies tends to be more continuous, spanning ages from a few $10$s Myr up to $10$ Gyr. The above mentioned gap could be explained by the following reasoning. Firstly, as mentioned above, NGC~7318A is considered an elliptical galaxy so we would not expect it to contribute many younger clusters. In any case, the SCCs which are just under $2$ Gyr could have faded below the detection limit at such a distance. For a given mass, they are fainter than the young star clusters ($\lesssim$ a few tens Myr) by $\sim 3$ mags and even a small amount of extinction may either push them below our detection limit or redden them, and so they will appear as older clusters ($> 2$ Gyr). Thirdly, previous work (\eg~\altcite{moles97}) has found that NGC~7318B is interacting with SQ for the first time. They draw this conclusion mainly from two observations: the $v_{\rm R}$ of NGC~7318B with respect to SQ ($\Delta v \sim 900$\kms) and its mostly unperturbed spiral structure, a sign that this galaxy has not experienced many interactions in its recent history. Thus, considering that galaxy interactions usually trigger star formation, without recent interactions with other galaxies, NGC~7318B will be expected to have a small number of SCCs younger than $2$ Gyr above our luminosity limit, consistent with a galaxy with continuous star formation typical of an isolated spiral. The gap can therefore be used to age-date the interaction from the onset of enhanced star formation.

	\subsection{NGC~7319}
The analysis of Figure~\ref{fig:N7319} shows that star formation in NGC~7319 has been going on throughout its history, reflected in the continuous distribution of SCCs along the evolutionary track. However, some peculiarities are observed. For example, the fraction of SCCs younger than $100$ Myr is lower than found in NGC~7318A/B and NSBR. Our analysis of Figure \ref{fig:mass.v.age} (which gives us the approximate values of the masses and ages of SCCs above the 50\% completeness level) shows that we should be able to detect a $100$ Myr old clusters down to $10^{4.0}$\Msun, relatively modest-sized clusters that are massive enough  to have a fully sampled initial mass function and therefore 'normal' colors (as investigated by \altcite{Silva-Villa2011}). Thus, the low number of $\lesssim 100$ Myr old clusters is likely to be real. Another observable is a somewhat tighter concentration of SCCs corresponding to the ages between approximately $100$ and $500$ Myr. In light of the models of interaction found in the literature, our results are consistent with the following scenario: the first traceable interaction for NGC~7319 was with NGC\emph{}~7320C and it happened $\sim 500$ Myr ago. It was the cause of the initial increase in the star formation rate. At the same time, this interaction stripped a large quantity of the gas from NGC~7319 and deposited it into the intergalactic medium (IGM), between galaxies NGC~7319 and NGC~7318A (\altcite{moles98}). Since the mass of the H\one\ gas in that region is estimated to be $1.4\times10^{10}h^{-1}_{100}$ \Msun\ (\altcite{moles97}), more than what would be expected in a galaxy with the mass of NGC~7319 ($3.2 \times 10^{11}$~\Msun; \altcite{renaud10}), gas from other galaxies may also contribute. The most likely candidates are NGC~7320C, which has no detectable H\one\ (\altcite{sulentic01}), and NGC~7318A. The models by \citet{renaud10} suggest that approximately $200$ Myr ago NGC~7319 had another interaction, that time with NGC~7318A (see \S \ref{sec:YT}), triggering a second wave of star formation and depleting the remaining galactic gas reservoir. At present, there is no neutral hydrogen present in NGC~7319 (\altcite{sulentic01}), which is consistent with the lack of evidence for recent ($\lesssim 20$ Myr) star formation as can be seen in Figure~\ref{fig:N7319}. 

	\subsection{Southern Debris Region}
The Southern Debris Region (SDR) is located between the NGC~7317 and NGC~7318A galaxies. It hosts low surface brightness structures, with sizes between 2 to 4 kpc, that might qualify as tidal-dwarf galaxies. As can be observed in Figure~\ref{fig:N7317_SDR}, SCCs in the SDR are divided into two groups, the old star clusters ($\gtrsim 2$ Gyr) and cluster candidates with ages between $50$ and $\sim 500$ Myr. The high number of old star clusters can be explained by the proximity of elliptical galaxy NGC~7317, since the expected radius of $\sim 40$ kpc for its old globular cluster system would overlap with the SDR. The radial extent of the GC system was calculated with the recipe from \citet{rhode07}, which relates the mass of a galaxy (derived from the mass-to-light ratio specific to that type of galaxies given its \vb -band luminosity) to the radial extent of the GC system. 

However, the origin of the Southern Debris region is not yet clear. On the one hand, the H\one\ in the SDR has the same velocity (in the range of $5710$---$5770$ \kms; \altcite{williams02}) as the shock front of NGC~7318B (\altcite{sulentic01}), suggesting that this region might be related to NGC~7318B. On the other hand, the age distribution of younger SCCs is consistent with the ages of the Young Tail. In \citet{renaud10}, the preferred model for the formation of the Young Tail is the interaction of NGC~7319 and NGC~7318A (\S~\ref{sec:YT}). Through that interaction NGC~7319 and NGC~7318A both produce tidal tails facing in opposite directions from each other: the tail of NGC  7319 points to the east, while the other one points west. The Southern Debris Region, with its group of younger star clusters with the ages $\lesssim 400$ Myr, could fit into this scenario as a part of the west pointing tail. However, there is a caveat. In order for this model to be plausible prior to the interaction, NGC~7318A is assumed to have been a spiral galaxy with a cold gaseous disk, which is still questionable as mentioned in \S \ref{sec:AB}. 

	\subsection{Northern Star Burst Region}
As previously mentioned, NGC~7318B is undergoing a collision with NGC~7318A and the IGM, deposited here by previous interactions involving NGC~7319, NGC~7320C, and perhaps NGC~7318A. This resulted in the formation of numerous star clusters, of which we were able to detect 110, a rather large number given this region is located outside of NGC~7318A \& B. The most noticeable features of this region are two tidal tails observed to the north of NGC~7318A/B that seemingly cross each other. \citet{renaud10} suspect that those tidal tails are physically overlapping, triggering intense star formation at the intersection. This is supported by the discoveries of strong H$_{\alpha}$, IR (\altcite{xu99}), and UV (\altcite{xu05}) emission at that location.

The hypothesis of the intersecting tidal tails is also supported by the H\one\ maps of SQ presented by \citet{williams02} who measure a velocity of $\sim 5950$ \kms for the H\one\ at the part of the tail closest to NGC 7318A. The velocity is increasing with the distance from NGC~7318A, reaching $\sim 6000$ \kms\ at the point of intersection of the two tails, and then continuing to increase along the eastern tail towards NGC~7318B up to to a value of $v \approx 6060$ \kms. The velocity of H\one\ in the intersection region ($\sim 6000$ \kms) is consistent with its being situated between NGC~7318A ($6671$ \kms) and NGC~7318B ($5766$ \kms). That, in turn, suggests that the two tidal tails are a product of the interaction between those galaxies.

From Figure~\ref{fig:colours}, we can see that the NSBR region has many young SCCs. Also, the gap in color-color plots between the intermediate-aged clusters (few Gyr) and younger clusters (few $100$ Myr) seen in other regions of SQ (\eg\ NGC~7318A/B, SDR) is not as pronounced here. Most probably, this is due to reddened young clusters filling any gaps, if present. Overall, the age estimates for this region are somewhat uncertain due to the presence of significant amounts of dust and gas. However, based on Figure~\ref{fig:NSBR} we 
see that the NSBR is a region with ongoing active star formation as we can observe a number of star clusters as young as a few Myr, consistent with G01.

%=============== SECTION 4 ===============
\section{History of interactions in SQ}\label{sec:history}

At the present moment, we cannot say anything certain about interactions that might have happened earlier ($>$ 1 Gyr ago) in the history of SQ.  However, based on the diffuse halo that surrounds SQ's core and NGC~7317, some authors (\eg\ \altcite{moles97}) suggest that interactions did occur prior to those that we discuss. 

In this section, we integrate results from the literature with the new information provided by our age-estimates of star clusters, whose formation we assume is triggered by dynamical interactions between galaxies.  We therefore advance the following timetable for the history of interactions between the members of SQ in the last Gyr.

The oldest event that has left a significant trace in the SQ system was a gravitational encounter between NGC~7319 with NGC~7320C (\altcite{moles97}; \altcite{sulentic01}) which caused the formation of the Old Tail.  Our analysis of the color-color diagram of star clusters detected in the Old Tail indicate this event happened approximately $400$ Myr ago.  Moreover, this encounter is also responsible for stripping the ISM from NGC~7319 and depositing it to the west of the galaxy (\altcite{moles98}).  Since NGC~7320C has no detectable H\one\ (\altcite{sulentic01}) and assuming that there was only one interaction between those galaxies (see \S~\ref{sec:OT} and \S~\ref{sec:YT}), we conclude that this collision also stripped all gas from NGC~7320C.

\begin{figure}
\begin{center}
    \subfigure{\includegraphics[width=1.0\textwidth]{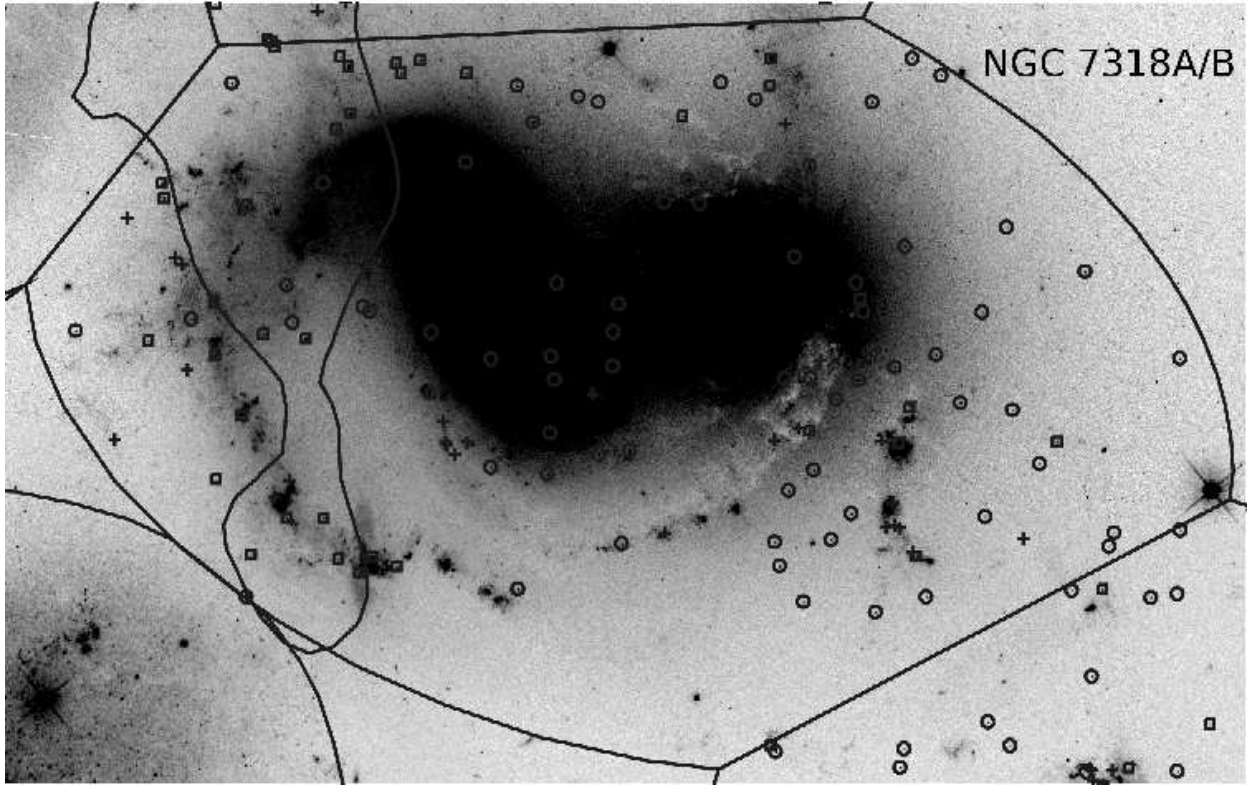}}\\
    \subfigure{\includegraphics[scale=0.75]{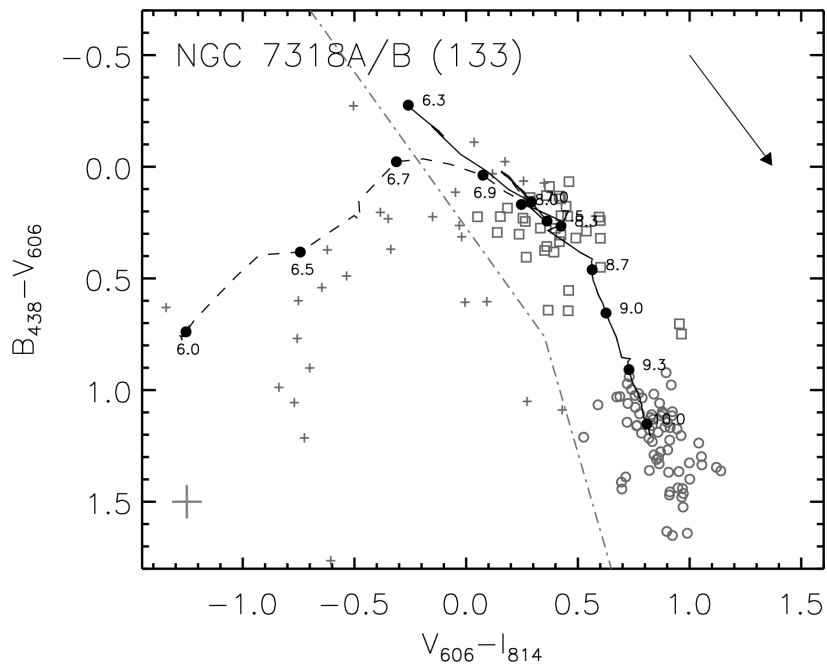}}
\end{center}
    \vspace{-0.75cm} 
    \caption{\captionfonts {\bf Top}: The spatial distribution of SCCs in the NGC~7318A/B region marked according to their ages (symbol definitions follow Figure \ref{fig:N7317_SDR}). {\bf Bottom}: Color-color plots of the same region. A lack of SCCs between the older ($\gtrsim 2$ Gyr) and younger ($\lesssim 500$ Myr) populations of SCCs is observed. Five out of the seven objects that are located in the box of $0.0$ mag $<$ \vb $-$ \ib\ $<1.1$ mag and $0.5$ mag $<$ \bb $-$ \vb\ $<0.8$ mag very likely to be reddened young clusters (see \S \ref{sec:AB}). However, this gap is somewhat unusual as recent studies (\eg~\altcite{gall10}, \altcite{konst10}) have shown that the distribution of SCCs for spiral galaxies tends to be more continuous, spanning ages from few $10$s Myr up to $10$ Gyr.}
\label{fig:N7318AB}
\end{figure}

\begin{figure}
\begin{center}
    \subfigure{\includegraphics[width=1.0\textwidth]{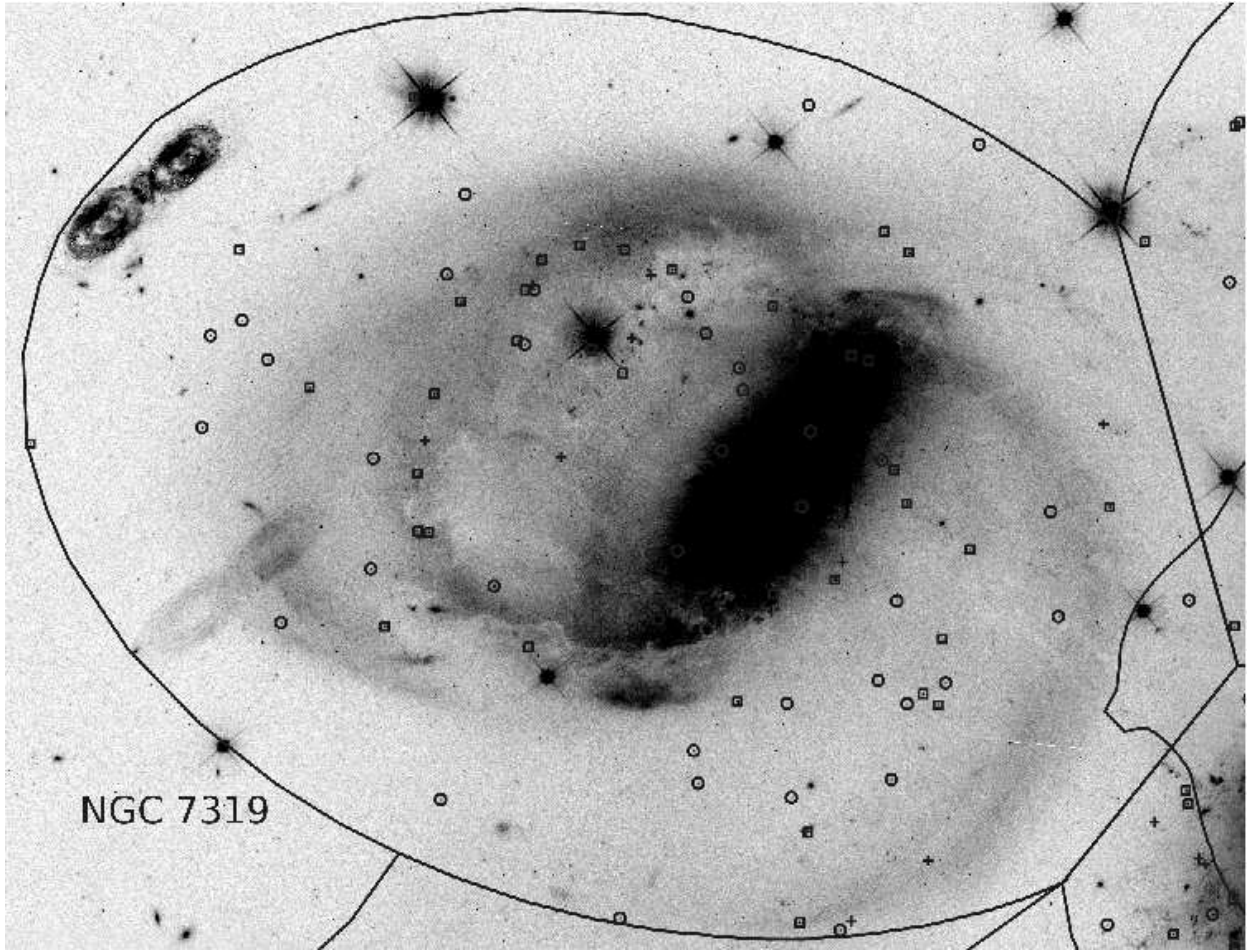}}\\
    \subfigure{\includegraphics[scale=0.75]{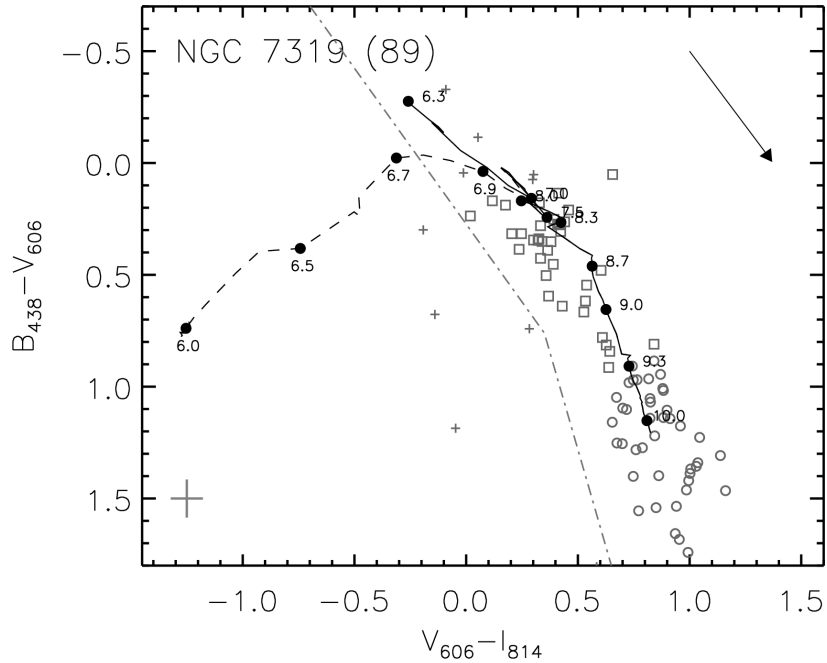}}
\end{center}
   \vspace{-0.75cm} 
   \caption{\captionfonts {\bf Top}: The spatial distribution of SCCs in the NGC~7319 region marked according to their ages (symbol definitions follow Figure \ref{fig:N7317_SDR}). {\bf Bottom}: Color-color plots of the same region. There are a number of intermediate-aged clusters along the northeast spiral arm. The number of SCCs younger than $100$ Myr is lower in comparison to NGC~7318A/B and NSBR regions. It would appear that star formation in this region was truncated $\sim 20$ Myr ago consistent with the observed lack of neutral hydrogen in NGC~7319.}
  \label{fig:N7319}
\end{figure}

\begin{figure}
\begin{center}
    \subfigure{\includegraphics[width=0.80\textwidth]{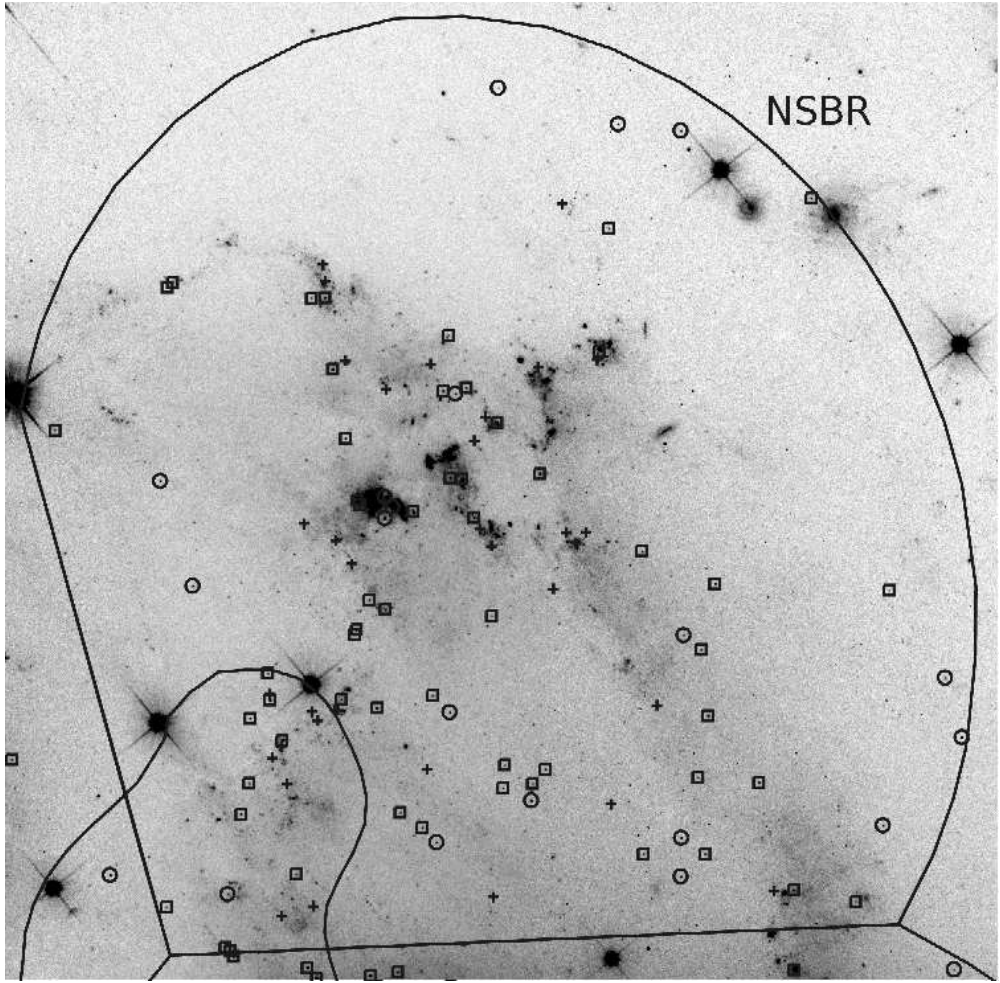}}\\
    \subfigure{\includegraphics[scale=0.75]{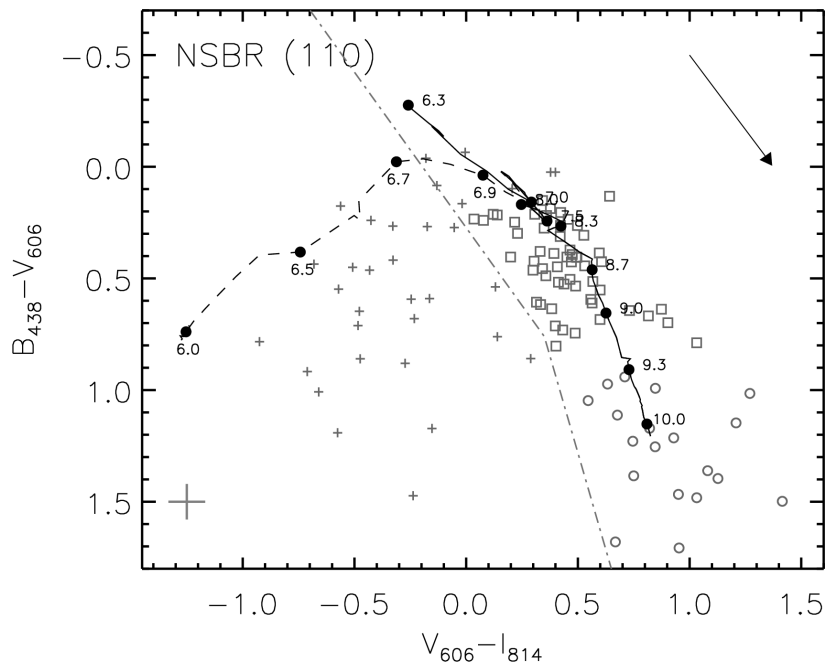}}
\end{center}
   \vspace{-0.75cm}   
   \caption{\captionfonts {\bf Top}: The spatial distribution of SCCs in the NSBR region marked according to their ages (symbol definitions follow Figure \ref{fig:N7317_SDR}). {\bf Bottom}: Color-color plots of the same region. A large number of very young SCCs ($< 10$ Myr) is observed. A gap in color-color plots between clusters of a few Gyr and a few $100$ Myr seen in other regions of SQ (\eg\ NGC~7318A/B, SDR) is not as prominent here, likely from reddened young clusters populating this region.}
\label{fig:NSBR}
\end{figure}

\begin{figure}
\begin{center}
   \includegraphics[width=1.0\textwidth]{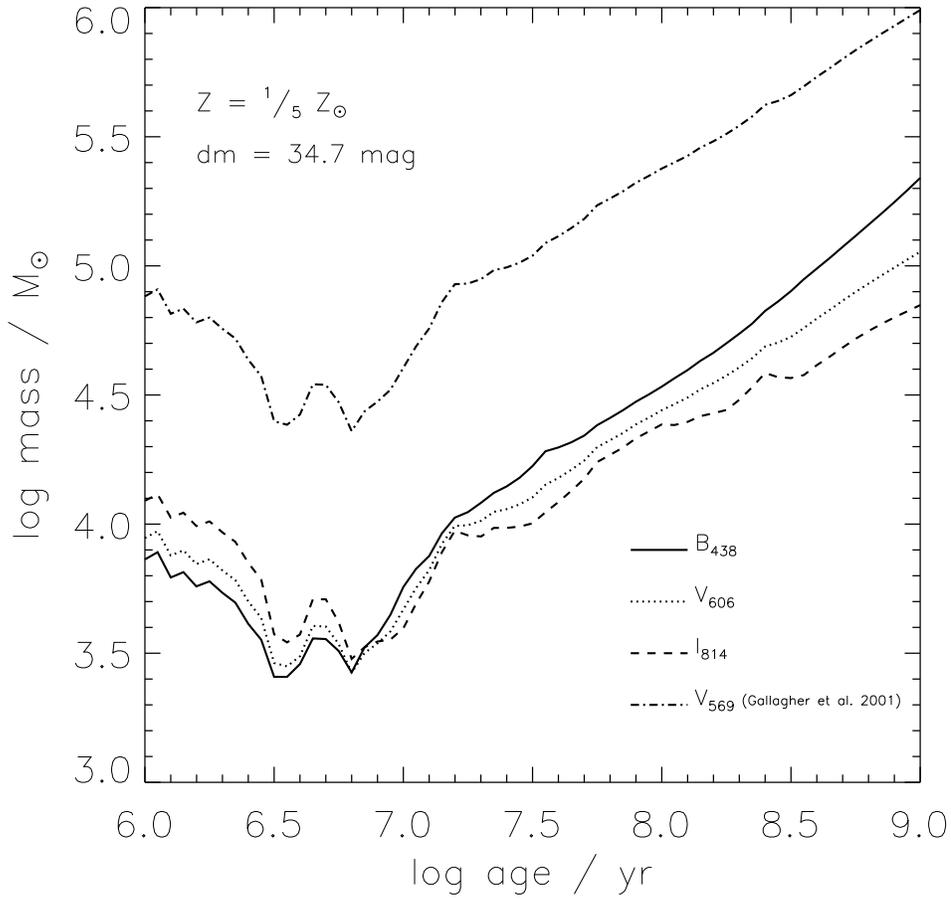}
\end{center}   
   \vspace{-0.5cm} 
   \caption{\captionfonts Plot of predicted cluster masses over time for 90\% completeness limits in the \bb, \vb, and \ib\ bands. For the intermediate-aged star clusters ($7<\log(t)<8$) we are capable of detecting star clusters with masses as low as $\sim 10^{4.5}$ \Msun. For comparison, the dashed-dotted line represents the estimated cluster masses detectable in \citet{Gallagher2001} based on 90\% completeness limit in the $V_{569}$-band.}
\label{fig:mass.v.age}
\end{figure}

Next, approximately $200$---$250$ Myr ago, based on the age estimates obtained in this paper, NGC~7319 experienced a close approach by NGC~7318A coming from the northeast (\altcite{renaud10}).  If we assume that NGC~7318A used to be a spiral galaxy (as assumed in \altcite{renaud10}), then the results of the interaction with NGC~7319 were pronounced. NGC~7319 formed a second tidal tail (the Young Tail; \altcite{xu05}), and star formation was triggered in both NGC 7319 itself and in the tail. This is seen in the  color-color plots of Figures~\ref{fig:YT_OT} and \ref{fig:N7318AB} as an over-density of star clusters corresponding to those ages. Specifically, the bulk of the younger star clusters in the Young Tail were formed at approximately the same time ($\sim 200$ Myr ago).  The low surface brightness features located south of NGC~7318A/B, and that currently do not belong to any of the earlier defined regions, might represent parts of the stripped spiral arms of NGC~7318A.  We infer this by incorporating the hydrodynamic modelling of \citet{renaud10} with our age-estimates for the young clusters in the SDR.  Most of the ISM from NGC~7318A and NGC~7319, along with the ISM from NGC~7320C, is then located between NGC~7318A and NGC~7319 (\altcite{renaud10}), in what becomes the NSBR.

More recently, the collision of NGC~7318B with the core occurred (\altcite{sulentic01}, \altcite{xu03}).  According to \citet{renaud10}, the collision was twofold; first NGC~7318B interacted with NGC~7318A, and then it collided with the ISM.  It is very difficult to determine the approximate time of the collision between NGC~7318B and NGC~7318A since our results from BVI data are distorted by the presence of gas and dust in that region (i.e., making SCCs likely to appear  older than they are).  Nevertheless, a crude estimate can be obtained from the color-color plot for the NGC~7318A/B region (Figure~\ref{fig:N7318AB}).  Considering the gap in the distribution of SCCs  discussed in \S~\ref{sec:AB}) and the increase in star cluster density in the vicinity of $\log(t)=8.4$---$8.5$ yr, we conclude that the upper limit on age of the first event is $\sim 250$---$300$ Myr.  In all probability, the real age of the event is younger since the NGC~7318B collision with the SQ's core would have likely happened after the formation of the Young Tail ($\sim 200$ Myr), purely by a  kinematical argument.  Assuming the radial velocity difference between the SQ core and NGC~7318B ($\sim 900$ \kms) to be constant for the last $250$ Myr, and the observational fact that NGC~7318B is in the midst of the collision with the IGM (assumed to be in the same plane with NGC~7319 and NGC~7318A), it would imply that NGC~7318B started at a distance of more than $230$ kpc (along the line-of-sight) from the present position.  This means that $250$ Myr ago NGC~7318B and NGC~7318A were too far apart for an interaction between them to be responsible for the increase of the star cluster numbers we observe. Additionally, NGC~7318B still possesses its spiral structure, another argument for a recent collision.  At the same time, based on Figures~\ref{fig:N7318AB} and \ref{fig:NSBR}, we can state with certainty that the collision of NGC~7318B with the core is still ongoing because we can detect a number of very young star clusters $< 10$ Myr old.

The future of the group is strongly dependent on the real spatial velocity dispersion between its members. However, based on the $v_{\rm R}$, some predictions can be made. Since NGC~7319 and NGC~7318A have almost identical radial velocities (see Table \ref{tab:res}) and close spatial proximity it is reasonable to conclude that these galaxies will eventually merge (\altcite{renaud10}). No clear hypothesis can be drawn for the future of the rest of the galaxies. NGC~7320C is relatively far from the group (at $\sim 85$ kpc in projection), NGC~7318B has a very large velocity difference with regards to the rest of the group, and we do not know the real trajectory of NGC~7317. It is possible that if NGC~7317 merges with NGC~7319 and NGC~7318A, resulting in a deeper gravitational well, NGC~7320C and/or NGC~7318B might not be able to escape.

%=============== SECTION 5 ===============
\section{Summary}\label{sec:summary}
Based on the sensitive, high S/N photometry that was obtained from ERS with WFC3 on \hst\ we were able to make robust age estimates of recent events in the history of SQ by studying the star cluster populations. Thus, we found the following:
\begin{itemize}
\item The star cluster population of NGC~7317 consists of clusters that are over $2$ Gyr old, indicating no recent epochs of significant star formation. From their luminosities, the majority of them can be considered globular clusters.

\item The Old Tail formed approximately $400$ Myr ago, based on the distribution of ages among its star clusters. The remarkably tight distribution and relatively young age of star clusters suggests in situ star formation, meaning that the Old Tail was formed with a large amount of gas.

\item Most of the star cluster candidates detected in the Young Tail are in the age range between $150$ and $200$ Myr suggesting this tidal feature formed $\lesssim 200$ Myr ago. 

\item NGC~7319 exhibits continuous star formation throughout its history, albeit at a lower rate over the past few tens of Myr presumably because of gas stripping from the most recent interaction. The SCC color distribution shows an over-density around $200$ Myr and a possible clump between $500$ Myr and $1$ Gyr. The first feature corresponds to the time of formation of the Young Tail, and the second feature could correspond to the time of formation of the Old Tail.

\item NGC~7318 A/B shows ongoing active star formation; there are a number of star clusters that are younger than $10$ Myr. This region also features a pronounced gap in the SCCs distribution in color space between $\sim 400$ Myr and $2$ Gyr, thus dating the onset of the interaction-induced star formation to $\lesssim 400$ Myr.

\item We have detected 110 SCCs in the Northern Star Burst region, a large number given the inter-galactic location of this region. The age estimates for this region are somewhat uncertain due to the presence of significant columns of dust and gas. However, we can conclude that star formation is ongoing in this region, given the large number of clusters younger than $10$ Myr.

\item The star cluster population of the Southern Debris region consists of two age groups: SCCs with ages between $50$ and $\sim 500$ Myr,dating perhaps to the interaction between NGC~7318A and NGC~7319, and $\gtrsim 2$ Gyr old clusters, likely stripped from NGC~7318A. It hosts low surface brightness structures, with sizes between 2 and 4 kpc, which might qualify as tidal-dwarf galaxies.

\end{itemize}

The combination of results obtained through our new optical photometry and empirical results and dynamical modelling in the literature has advanced our understanding of the history of interactions in this fascinating group of galaxies. However, there are still some unanswered questions. For example, the morphological type of NGC~7318A is still a mystery. Although some of the observed properties correspond well with it being an elliptical galaxy, the presence of tidal arms as well as regions with debris of low surface brightness would rather suggest that this galaxy was a spiral not that long ago. It is possible that NGC~7318A is in fact the stripped core of a spiral galaxy.

Another unclear point in SQ's history is the formation of the SDR region. The present images obtained with the ERS are not deep enough to see if there are some low surface brightness structures that might be associated with the SDR. The possibility that we adopted in this paper is that this region originated at the same time as the Young Tail and is a direct result of the interaction between NGC~7319 and NGC~7318A. The same scenario is proposed by \citet{renaud10}. 

Additional high resolution photometry in the $U$-band would improve the age-estimates of clusters in the $10$---$500$ Myr age range, and thus tighten the constraints on the time scale of interaction-induced star formation, particularly along the X-ray shock region.

%-----------------------------------------------------------------------------------
\vspace{2cm}
We thank Florent Renaud for his useful feedback. Support for this work was provided by the National Science and Engineering Research Council of Canada and the Ontario Early Researcher Award (KF \& SCG). SCG also thanks the Rotman Institute for its hospitality during
Summer 2010. R. C. is grateful for support from NSF through CAREER award 0847467. J. C. C. and I. S. K. acknowledge their support by the NSF through award 0908984. This research has made use of SAOImage DS9, developed by Smithsonian Astrophysical Observatory, and of the NASA/IPAC Extragalactic Database (NED) which is operated by the Jet Propulsion Laboratory, California Institute of Technology, under contract with the National Aeronautics and Space Administration. 

\renewcommand{\baselinestretch}{1.0}

\bibliographystyle{apj}
\bibliography{MyCollection}

\begin{thebibliography}{39}
\expandafter\ifx\csname natexlab\endcsname\relax\def\natexlab#1{#1}\fi

\bibitem[{Allen \& {Sullivan III}(1980)}]{allen80}
Allen, R.~J., \& {Sullivan III}, W.~T. 1980, Astronomy and Astrophysics, 84,
  181

\bibitem[{Barmby {et~al.}(2006)Barmby, Kuntz, Huchra, \& Brodie}]{barmby06}
Barmby, P., Kuntz, K.~D., Huchra, J.~P., \& Brodie, J.~P. 2006, The
  Astronomical Journal, 132, 883

\bibitem[{Bergvall {et~al.}(2003)Bergvall, Laurikainen, \& Aalto}]{bergvall03}
Bergvall, N., Laurikainen, E., \& Aalto, S. 2003, Astronomy and Astrophysics,
  405, 31

\bibitem[{Conti {et~al.}(1996)Conti, Leitherer, \& Vacca}]{conti96}
Conti, P.~S., Leitherer, C., \& Vacca, W.~D. 1996, The Astrophysical Journal,
  461, L87+

\bibitem[{de~Grijs {et~al.}(2003)de~Grijs, Anders, Bastian, Lynds, Lamers, \&
  O'Neil}]{degrijs03}
de~Grijs, R., Anders, P., Bastian, N., Lynds, R., Lamers, H. J. G. L.~M., \&
  O'Neil, E.~J. 2003, Monthly Notices of the Royal Astronomical Society, 343,
  1285

\bibitem[{de~Vaucouleurs {et~al.}(1991)de~Vaucouleurs, de~Vaucouleurs, {Corwin
  Jr.}, Buta, Paturel, \& Fouqu\'{e}}]{rc3}
de~Vaucouleurs, G., de~Vaucouleurs, A., {Corwin Jr.}, H.~G., Buta, R.~J.,
  Paturel, G., \& Fouqu\'{e}, P. 1991, {Third Reference Catalogue of Bright
  Galaxies. Volume I: Explanations and references. Volume II: Data for galaxies
  between 0$^{h}$ and 12$^{h}$. Volume III: Data for galaxies between 12$^{h}$
  and 24$^{h}$.} (New York, NY: Springer)

\bibitem[{Falco {et~al.}(1999)Falco, Kurtz, Geller, Huchra, Peters, Berlind,
  Mink, Tokarz, \& Elwell}]{UZC}
Falco, E.~E., {et~al.} 1999, Publications of the Astronomical Society of the
  Pacific, 111, 438

\bibitem[{Gallagher {et~al.}(2001)Gallagher, Charlton, Hunsberger, Zaritsky, \&
  Whitmore}]{Gallagher2001}
Gallagher, S.~C., Charlton, J.~C., Hunsberger, S.~D., Zaritsky, D., \&
  Whitmore, B.~C. 2001, The Astronomical Journal, 122, 163

\bibitem[{Gallagher {et~al.}(2010)Gallagher, Durrell, Elmegreen, Chandar,
  English, Charlton, Gronwall, Young, Tzanavaris, Johnson, {Mendes de
  Oliveira}, Whitmore, Hornschemeier, Maybhate, \& Zabludoff}]{gall10}
Gallagher, S.~C., {et~al.} 2010, The Astronomical Journal, 139, 545

\bibitem[{Gutierrez {et~al.}(2002)Gutierrez, Lopez‐Corredoira, Prada, \&
  Eliche}]{Gutierrez2002}
Gutierrez, C.~M., Lopez‐Corredoira, M., Prada, F., \& Eliche, M.~C. 2002, The
  Astrophysical Journal, 579, 592

\bibitem[{Hickson(1982)}]{Hickson1982}
Hickson, P. 1982, The Astrophysical Journal, 255, 382

\bibitem[{Konstantopoulos {et~al.}(2010)Konstantopoulos, Gallagher, Fedotov,
  Durrell, Heiderman, Elmegreen, Charlton, Hibbard, Tzanavaris, Chandar,
  Johnson, Maybhate, Zabludoff, Gronwall, Szathmary, Hornschemeier, English,
  Whitmore, {Mendes de Oliveira}, \& Mulchaey}]{konst10}
Konstantopoulos, I.~S., {et~al.} 2010, The Astrophysical Journal, 723, 197

\bibitem[{Kroupa(1998)}]{kroupa98}
Kroupa, P. 1998, in Astronomical Society of the Pacific Conference Series, Vol.
  134, Brown Dwarfs and Extrasolar Planets, ed. .~M. R. Z.~O. {R. Rebolo, E. L.
  Martin}, 483

\bibitem[{Larsen(1999)}]{larsen99}
Larsen, S. r.~S. 1999, Astronomy and Astrophysics Supplement Series, 139, 393

\bibitem[{Leitherer {et~al.}(1999)Leitherer, Schaerer, Goldader, Delgado,
  Robert, Kune, de~Mello, Devost, \& Heckman}]{leith99}
Leitherer, C., {et~al.} 1999, The Astrophysical Journal Supplement Series, 123,
  3

\bibitem[{Marigo {et~al.}(2008)Marigo, Girardi, Bressan, Groenewegen, Silva, \&
  Granato}]{marigo08}
Marigo, P., Girardi, L., Bressan, A., Groenewegen, M. A.~T., Silva, L., \&
  Granato, G.~L. 2008, Astronomy and Astrophysics, 482, 883

\bibitem[{{Mendes de Oliveira} \& Hickson(1994)}]{mendes94}
{Mendes de Oliveira}, C., \& Hickson, P. 1994, The Astrophysical Journal, 427,
  684

\bibitem[{Moles {et~al.}(1998)Moles, Marquez, \& Sulentic}]{moles98}
Moles, M., Marquez, I., \& Sulentic, J.~W. 1998, Astronomy \& Astrophysics,
  334, 8

\bibitem[{Moles {et~al.}(1997)Moles, Sulentic, \& M\'{a}rquez}]{moles97}
Moles, M., Sulentic, J.~W., \& M\'{a}rquez, I. 1997, The Astrophysical Journal,
  485, L69

\bibitem[{Nilson(1973)}]{nilson73}
Nilson, P. 1973, Nova Acta Regiae Soc. Sci. Upsaliensis Ser. V

\bibitem[{Rejkuba {et~al.}(2005)Rejkuba, Greggio, Harris, Harris, \&
  Peng}]{rejkuba05}
Rejkuba, M., Greggio, L., Harris, W.~E., Harris, G. L.~H., \& Peng, E.~W. 2005,
  The Astrophysical Journal, 631, 262

\bibitem[{Renaud {et~al.}(2010)Renaud, Appleton, \& Xu}]{renaud10}
Renaud, F., Appleton, P.~N., \& Xu, C.~K. 2010, The Astrophysical Journal, 724,
  80

\bibitem[{Rhode {et~al.}(2007)Rhode, Zepf, Kundu, \& Larner}]{rhode07}
Rhode, K.~L., Zepf, S.~E., Kundu, A., \& Larner, A.~N. 2007, The Astronomical
  Journal, 134, 1403

\bibitem[{Robin {et~al.}(2003)Robin, Reyl\'{e}, Derri\`{e}re, \&
  Picaud}]{robin03}
Robin, A.~C., Reyl\'{e}, C., Derri\`{e}re, S., \& Picaud, S. 2003, Astronomy
  and Astrophysics, 409, 523

\bibitem[{Salzer {et~al.}(2005)Salzer, Jangren, Gronwall, Werk, Chomiuk,
  Caperton, Melbourne, \& McKinstry}]{salzer05}
Salzer, J.~J., Jangren, A., Gronwall, C., Werk, J.~K., Chomiuk, L.~B.,
  Caperton, K.~A., Melbourne, J., \& McKinstry, K. 2005, The Astronomical
  Journal, 130, 2584

\bibitem[{Saracco \& Ciliegi(1995)}]{sqmetal}
Saracco, P., \& Ciliegi, P. 1995, \aap, 301, 348

\bibitem[{Scheepmaker {et~al.}(2007)Scheepmaker, Haas, Gieles, Bastian, Larsen,
  \& Lamers}]{scheep07}
Scheepmaker, R.~A., Haas, M.~R., Gieles, M., Bastian, N., Larsen, S. r.~S., \&
  Lamers, H. J. G. L.~M. 2007, Astronomy and Astrophysics, 469, 925

\bibitem[{Schlegel {et~al.}(1998)Schlegel, Finkbeiner, \& Davis}]{schlegel98}
Schlegel, D.~J., Finkbeiner, D.~P., \& Davis, M. 1998, The Astrophysical
  Journal, 500, 525

\bibitem[{Shostak {et~al.}(1984)Shostak, Allen, \& {Sullivan III}}]{shostak84}
Shostak, G.~S., Allen, R.~J., \& {Sullivan III}, W.~T. 1984, Astronomy and
  Astrophysics, 139, 15

\bibitem[{Silva-Villa \& Larsen(2011)}]{Silva-Villa2011}
Silva-Villa, E., \& Larsen, S.~S. 2011, Astronomy \& Astrophysics, 529, A25

\bibitem[{Stephan(1877)}]{steph1877}
Stephan, M. 1877, Monthly Notices of the Royal Astronomical Society, 37

\bibitem[{Stetson(1987)}]{stetson87}
Stetson, P.~B. 1987, Publications of the Astronomical Society of the Pacific,
  99, 191

\bibitem[{Sulentic {et~al.}(2001)Sulentic, Rosado, Dultzin-Hacyan,
  Verdes-Montenegro, Trinchieri, Xu, \& Pietsch}]{sulentic01}
Sulentic, J.~W., Rosado, M., Dultzin-Hacyan, D., Verdes-Montenegro, L.,
  Trinchieri, G., Xu, C., \& Pietsch, W. 2001, The Astronomical Journal, 122,
  2993

\bibitem[{Vacca \& Conti(1992)}]{vacca92}
Vacca, W.~D., \& Conti, P.~S. 1992, The Astrophysical Journal, 401, 543

\bibitem[{Whitmore {et~al.}(1999)Whitmore, Zhang, Leitherer, Fall, Schweizer,
  \& Miller}]{whitmore99}
Whitmore, B.~C., Zhang, Q., Leitherer, C., Fall, S.~M., Schweizer, F., \&
  Miller, B.~W. 1999, The Astronomical Journal, 118, 1551

\bibitem[{Williams {et~al.}(2002)Williams, Yun, \&
  Verdes-Montenegro}]{williams02}
Williams, B.~A., Yun, M.~S., \& Verdes-Montenegro, L. 2002, The Astronomical
  Journal, 123, 2417

\bibitem[{Xu {et~al.}(1999)Xu, Sulentic, \& Tuffs}]{xu99}
Xu, C., Sulentic, J.~W., \& Tuffs, R. 1999, The Astrophysical Journal, 512, 178

\bibitem[{Xu {et~al.}(2003)Xu, Lu, Condon, Dopita, \& Tuffs}]{xu03}
Xu, C.~K., Lu, N., Condon, J.~J., Dopita, M., \& Tuffs, R.~J. 2003, The
  Astrophysical Journal, 595, 665

\bibitem[{Xu {et~al.}(2005)Xu, Iglesias-P\'{a}ramo, Burgarella, Rich, Neff,
  Lauger, Barlow, Bianchi, Byun, Forster, Friedman, Heckman, Jelinsky, Lee,
  Madore, Malina, Martin, Milliard, Morrissey, Schiminovich, Siegmund, Small,
  Szalay, Welsh, \& Wyder}]{xu05}
Xu, C.~K., {et~al.} 2005, The Astrophysical Journal, 619, L95

\end{thebibliography}

%===============   TABLES   ===============

%%%--- Table 1
\newpage
\begin{table*}
	\center
    \caption{Basic information on HCG~92 (members \emph a through \emph e). \tablenotemark{a}} 
    
	\begin{tabular}{ccccccc} 
	\hline
	\hline
	Identifier		&Hickson 	& Coordinates	& Type & B 	& $v_R$\\
				&designation\tablenotemark{b}	& (J2000)		&	   & (mag)& (\kms)\\
	\hline
	NGC~7317	& 92e & 22~35~51.9 +33~56~41	& E1	& 15.3 & 6637\\
	NGC~7318A	& 92d & 22~35~56.8 +33~57~57	& E2(pec)& 14.4 & 6671\\
	NGC~7318B	& 92b & 22~35~58.4 +33~57~57	& Sbc 	& 14.9 & 5766\\
	NGC~7319	& 92c & 22~36~03.7 +33~58~35	& Sbc	& 14.8 & 6652\\
	NGC~7320 \emph{(foreground)}	& 92a & 22~36~03.4 +33~56~54 & Sd		& 13.8 & 739 \\
	\hline
	\label{tab:tab1} 	
	\end{tabular}
%	\vspace*{-1.2cm}
    \tablenotetext{a}{From The Updated Zwicky Catalog (UZC), \citet{UZC}}
    \tablenotetext{b}{\citet{Hickson1982}}
\end{table*}

%%%--- Table 2
\newpage
\begin{table*}[ht]
	\center
	\caption{Properties of Star Cluster and Globular Cluster Candidates.\tablenotemark{a}}
 	\begin{tabular}{cccccc} 
	\hline
	\hline
	Regions			& \multicolumn{2}{c}{\# of SCCs} \\
					& Total & With Nebular Emission  \\
	\hline
	NGC~7317					& 64   & 0   \\
	NGC~7319					& 89   & 5   \\
	NGC~7318A/B					& 133  & 22  \\
	Young Tail					& 25   & 1   \\	
	Old Tail					& 33   & 0   \\
	Northern Star Burst Region	& 110  & 28  \\
	Southern Debris Region		& 42   & 3   \\
	X-Ray Shock Front			& 41   & 10  \\
	\hline
	\label{tab:res} 	
	\end{tabular}
%	\vspace*{-1.0cm}
	\tablenotetext{a}{The number of Star Cluster Candidates (SCCs; $M_{V_{606}} < -9$) detected in SQ. The algorithm of detecting SCCs is described in \S \ref{sec:SCCs}. The SCCs are considered to have nebular emission if their position in the color-color plot (\eg\ Figure \ref{fig:colours}) is to the left of the broken dot-dashed green line.}
\end{table*}

%===============   FIGURES   ===============

%%%--- Figure: $-color picture of SQ & Finder (Region Definer)---%%%
  
 %%%--- Figure: Colour-Magnitude ---%%%  

%%%--- Figure: Colour-Colour & Colour-Magnitude ---%%%

%%%--- Figure: Colors by region ---%%%

%%%--- Figure: N7317_SDR ---%%%

%%%--- Figure: Break in 7320 SCCs spatial distribution ---%%%

%%%--- Figure: YT_OT ---%%%

%%%--- Figure: N7318AB ---%%%

%%%--- Figure: N7319 ---%%%

%%%--- Figure: NSBR ---%%%

%%--- Figure: mass.v.age ---%%%

\end{document}